\documentclass[twocolumn]{aastex701}
\usepackage{graphicx}
\usepackage{color}
\usepackage{amsmath}
\usepackage{mathtools}
\usepackage{tcolorbox}
\usepackage{listings}
\usepackage{enumitem}
\usepackage{textcomp}
\RequirePackage{snapshot}
\definecolor{codegreen}{rgb}{0,0.6,0}
\definecolor{codegray}{rgb}{0.3,0.3,0.3}
\definecolor{codepurple}{rgb}{0.58,0,0.82}
\definecolor{backcolour}{rgb}{0.9,0.9,0.9}

\lstdefinestyle{mystyle}{
    backgroundcolor=\color{backcolour},   
    commentstyle=\color{codegreen},
    keywordstyle=\color{magenta},
    numberstyle=\tiny\color{codegray},
    stringstyle=\color{codepurple},
    basicstyle=\ttfamily\footnotesize,
    breakatwhitespace=false,         
    breaklines=true,                 
    captionpos=t,                    
    keepspaces=true,                 
    numbers=left,                    
    numbersep=5pt,                  
    showspaces=false,                
    showstringspaces=false,
    showtabs=false,                  
    tabsize=2
}

\newcommand{\geotr}{\earth_r}

\newcommand{\vect}[1]{\boldsymbol{#1}}
\newcommand{\murel}{\mu_{\mathrm{rel},\sun}}

\newcommand{\murelgeotr}{\mu_{\mathrm{rel},\geotr}}
\newcommand{\murelgeotrE}{\mu_{\mathrm{rel},\geotr,E}}
\newcommand{\murelgeotrN}{\mu_{\mathrm{rel},\geotr,N}}
\newcommand{\murelvec}{\vect{\mu}_{\boldsymbol{\mathrm{rel}},\sun}}
\newcommand{\murelvecgeo}{\vect{\mu}_{\boldsymbol{\mathrm{rel}},\earth}}
\newcommand{\murelvecgeotr}{\vect{\mu}_{\boldsymbol{\mathrm{rel}},\geotr}}

\newcommand{\murelhat}{\boldsymbol{\hat{\mu}}_{\boldsymbol{\mathrm{rel}},\sun}}

\newcommand{\murelhatgeotr}{\boldsymbol{\hat{\mu}}_{\boldsymbol{\mathrm{rel}},\geotr}}

\newcommand{\tnot}{t_{0,\sun}}
\newcommand{\uvecohat}{\boldsymbol{\hat{u}}_{\boldsymbol{0},\sun}}

\newcommand{\uvec}{\vect{u}_{\sun}}
\newcommand{\uvecgeo}{\vect{u}_{\earth}}
\newcommand{\uvecgeotr}{\vect{u}_{\geotr}}
\newcommand{\uveco}{\vect{u}_{\boldsymbol{0},\sun}}

\newcommand{\uvecogeotr}{\vect{u}_{\boldsymbol{0},\geotr}}

\newcommand{\ugeotrE}{u_{\geotr,E}}
\newcommand{\ugeotrN}{u_{\geotr,N}}
\newcommand{\uoE}{u_{0,\sun,E}}
\newcommand{\uoN}{u_{0,\sun,N}}

\newcommand{\deltavec}{\vect{\delta}_{\boldsymbol{c}}}

\newcommand{\tE}{t_{E,\sun}}
\newcommand{\tEgeotr}{t_{E,\geotr}}
\newcommand{\thetaE}{\theta_E}
\newcommand{\uo}{u_{0,\sun}}
\newcommand{\uogeotr}{u_{0,\geotr}}

\newcommand{\piEvec}{\vect{\pi}_{E,\sun}}
\newcommand{\piEvecgeotr}{\vect{\pi}_{E,\geotr}}
\newcommand{\piE}{\pi_E}
\newcommand{\piEhat}{\hat{\boldsymbol{\pi}}_{E,\sun}}
\newcommand{\pirel}{\pi_{\mathrm{rel}}}

\newcommand{\tnotgeotr}{t_{0,\geotr}}

\newcommand{\piEN}{\pi_{E, N}}
\newcommand{\piEE}{\pi_{E, E}}

\newcommand{\x}{E}
\newcommand{\y}{N}

\newcommand{\musvec}{\vect{\mu}_{\boldsymbol{S},\sun}}
\newcommand{\mulvec}{\vect{\mu}_{\boldsymbol{L},\sun}}
\newcommand{\Xsovec}{\vect{X}_{\boldsymbol{S,0},\sun}}
\newcommand{\Xlovec}{\vect{X}_{\boldsymbol{L,0},\sun}}

\newcommand{\Xlvec}{\vect{X}_{\boldsymbol{L},\sun}}
\newcommand{\Xsvecgeo}{\vect{X}_{\boldsymbol{S},\earth}}
\newcommand{\Xlvecgeo}{\vect{X}_{\boldsymbol{L},\earth}}
\newcommand{\Xcentvec}{\vect{X}_{\boldsymbol{cent},\earth}}
\newcommand{\Pvec}{\vect{P}}
\newcommand{\Pdotvec}{\vect{\dot{P}}}
\newcommand{\PE}{P_E}
\newcommand{\PN}{P_N}
\newcommand{\PdotE}{\dot{P}_E}
\newcommand{\PdotN}{\dot{P}_N}
\newcommand{\bsff}{b_{sff}}
\newcommand{\magS}{mag_S}
\newcommand{\magcent}{mag_{cent}}
\newcommand{\msun}{M_\odot}
\newcommand{\berkeley}{\affiliation{Department of Astronomy, University of California, Berkeley, CA 94720, USA}}

\newcommand{\bagle}{BAGLE}

\begin{document}

\title{The BAGLE Python Package for Bayesian Analysis of Gravitational Lensing Events}
\shorttitle{BAGLE}
\shortauthors{Lu et al.}

\author[0000-0001-9611-0009]{J. R. Lu}
\berkeley
\email[show]{jlu.astro@berkeley.edu}

\author[0000-0002-7226-0659]{Michael Medford}
\affiliation{Independent Researcher}
\email{michaelmedford@gmail.com}

\author[0000-0002-6406-1924]{Casey Y. Lam}
\affiliation{Observatories of the Carnegie Institution for Science, Pasadena, CA 91101, USA}
\email{clam@carnegiescience.edu}

\author[0009-0002-6097-9030]{T. Dex Bhadra}
\affiliation{Department of Astronomy, University of Maryland, College Park, MD 20742, USA}
\affiliation{Code 667, NASA Goddard Space Flight Center, Greenbelt, MD 20771, USA}
\email{tmbhadra@umd.edu}

\author[0000-0003-4591-3201]{Macy J. Huston}
\berkeley
\email{mhuston@berkeley.edu}

\author[0000-0002-0287-3783]{Natasha S. Abrams}
\berkeley
\email{nsabrams@berkeley.edu}

\author[0000-0003-0652-1862]{Edward Broadberry} % finite source
\affiliation{Department of Astronomy, University of Maryland, College Park, MD 20742, USA}
\email{edbroad@umd.edu}

\author[0009-0001-3654-3088]{Jeff Chen}
\berkeley
\email{jechen00@berkeley.edu}

\author[0000-0002-5029-3257]{Sean K. Terry}
\affiliation{Department of Astronomy, University of Maryland, College Park, MD 20742, USA}
\affiliation{Code 667, NASA Goddard Space Flight Center, Greenbelt, MD 20771, USA}
\email{skterry@umd.edu}

\author[0000-0003-0547-8444]{Nijaid Arredondo}
\affiliation{Department of Physics,
University of Illinois at Urbana-Champaign, Urbana, Illinois 61801, USA}
\email{josena2@illinois.edu}

\author[0000-0002-1395-5426]{Andrew Scharf}
\berkeley
\email{andrew\_scharf@berkeley.edu}

\date{\today}

\begin{abstract}
We present the open-source Python package, BAGLE (Bayesian Analysis of Gravitational Lensing Events), which enables modeling and joint fitting of photometric and astrometric data sets. 
We describe the model parameterizations and present the equations for microlensing events containing either a point-source, point-lens or a finite-source, point-lens geometry both with and without microlensing parallax due to the motion of the Earth or a satellite around the Sun. Conversions between different coordinate reference frames are also derived. We compare our model light curves to those from other papers and microlens modeling software, finding good agreement, although with some differences in finite-source models at a $\sim$1\% level detectable with upcoming observations from space-based facilities. 
We also use \bagle~to demonstrate the impact of changing lens mass, lens distance, and blended source flux fraction on photometric lightcurves and astrometric trajectories in preparation for upcoming Gaia data releases and the launch of the Nancy Grace Roman Space Telescope and its Galactic Bulge Time Domain Survey (GBTDS). 
In particular, we show that Roman GBTDS will detect significant microlensing parallax signals for events that are 2$\times$ shorter in duration than from ground-based surveys. Additionally, long-duration events with durations of $\tE>100$ days will yield microlensing parallax uncertainties of $\sigma_{\pi_E}<0.01$ with Roman, enabling confident identification of isolated stellar-mass black holes that can be modeled both astrometrically and photometrically with BAGLE for precise mass determinations. 
BAGLE is an open-source code and community development is encouraged.
\end{abstract}

\section{Introduction}

Gravitational microlensing is a key physical phenomenon for detecting dark or low-luminosity objects such as exoplanets or black holes\citep{Einstein:1936,Refsdal:1964,Paczynski:1986,Alcock:1995}. Microlensing occurs when a background source star passes behind a foreground object with mass (e.g., a black hole), and the foreground object acts as a lens to amplify and distort the background starlight. Microlensing has been used to (1) detect low-mass and widely separated exoplanets, or even free floating planets \citep{Mao:1991,Bennett:1999,Gaudi:2002,Gaudi2012}, (2) discover the first isolated stellar-mass black hole \citep{Lam:2022, Sahu:2022, Mroz:2022, Lam:2023-OB110462, Sahu:2025}, (3) search for primordial black holes that may make up dark matter \citep{Alcock:2000,Tisserand:2007,Griest:2013,Mroz:2024} and (4) weigh white dwarfs (WDs) to test WD mass-radius relations  \citep{Sahu2017, McGill2023}. 

Modeling the transient photometric and astrometric signals from microlensing events is complex as there are parameter degeneracies and multi-modal solutions. Microlensing model frameworks were largely developed when only photometric microlensing was observationally detectable. Microlensing models like pyLIMA \citep{Bachelet2017}, VBMicrolensing \citep{Bozza2010,Bozza2018,Bozza2021,Bozza2024}, and MulensModel \citep{Poleski2019} have been widely successful in using photometric quantities to fit microlensing events.

We present BAGLE: Bayesian Analysis of Gravitational Lensing Events -- an open-source, fully Python package to model photometric and astrometric microlensing events. BAGLE's primary strength is that a full suite of modeling, fitting, and analysis tools are available for both astrometric and photometric data sets. 
BAGLE provides an implementation of the gravitational lensing theory extensively described in the literatures \citep[e.g.]{Refsdal:1964, Gould:1992, Alcock:1995, Hog:1995, Dominik:1998a, Mao:1998, Han:2000, Belokurov:2002, An:2002, Gould:2014, Nucita:2016, Bozza2021},
BAGLE supports single and binary sources and lenses, dark and luminous lenses, Earth and satellite parallax, point sources and finite-size sources, and 
binary orbital motion or simpler approximations (linear and acceleration). In this paper, we introduce single source and lens models. Binary models are presented in \citet{Bhadra:2025} and incorporation of Gaussian process noise is presented in \citet{Chen:2025}.

BAGLE is introduced in \S\ref{sec:bagle}.
The underlying microlensing model mathematics are described in detail in \S\ref{sec:models}. 
We take a pedantic approach as careful and comprehensive descriptions of reference frames, 
time standards, and coordinate conversions are critical for astrometric microlensing, in particular. 
Description of the extensible code architecture is also described in \S\ref{sec:bagle_implemenation}.
Next, \bagle's model-fitting approach is described in \S\ref{sec:bagle_fits}. Code examples are included throughout these sections. \S\ref{sec:validation} presents validation of \bagle~against other packages and \S\ref{sec:validate_fitter} validates the model fitting methods. 
Results using \bagle~to show the impact of lens mass, parallax, blending on both photometric and astrometric microlensing signals are presented in \S\ref{sec:results} along with runtime comparisons between \bagle~and other packages.
We conclude and present future directions in \S\ref{sec:conclusion}. 
Appendix \S\ref{sec:bagle_dev} presents how BAGLE can be extended to implement 
new model parameterizations or utilize other fitting packages. 

\section{BAGLE \label{sec:bagle}}

\begin{figure}
    \includegraphics[width=0.5\textwidth]{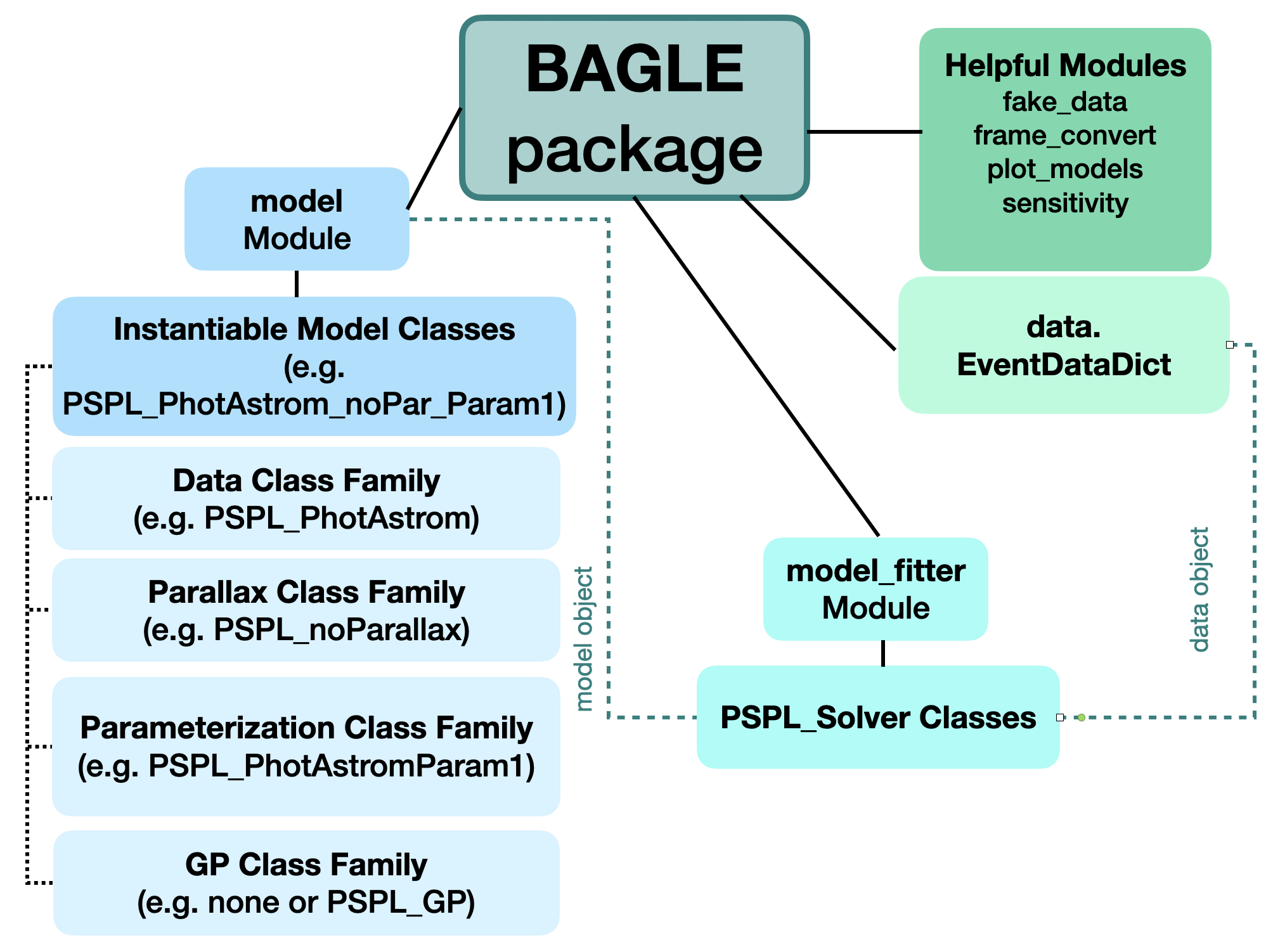}
    \caption{\label{fig:bagle}
    The BAGLE python package includes microlensing event models, data, model fitters, and numerous utility functions.
    }
\end{figure}

BAGLE is a Python package designed for modeling and fitting gravitational lensing events in the Milky Way or nearby Universe. 
The {\tt bagle} package includes modules for making models of individual events ({\tt bagle.model}), storing photometric and astrometric data ({\tt bagle.data}), fitting models to data ({\tt bagle.model\_fitter}), and numerous utilities for plotting and coordinate conversion as shown in Figure \ref{fig:bagle}.
BAGLE is fully open source (available on GitHub\footnote{\url{https://github.com/MovingUniverseLab/BAGLE_Microlensing}}) and is pip installable ({\tt pip install bagle}).
Full documentation is available via ReadTheDocs\footnote{\url{https://bagle.readthedocs.io/en/latest/}} along with examples of how to instantiate BAGLE models and run fits. 
This paper presents v1.0.2 of the code. 

% See figures.key for editable figure.

BAGLE has several dependencies that are required to be installed before running. Some typical libraries like \texttt{Astropy} \citep{astropy:2013} \texttt{Numpy} \citep{oliphant2006guide} and \texttt{Pytest} \citep{pytestx.y} are used for various astronomical functions, creating and manipulating arrays, and testing parts of the code. Further, the main fitting functions of BAGLE rely on \texttt{PyMultiNest} 
\citep{pymultinest} to perform nested sampling. Lastly, \texttt{celerite} \citep{celerite} is required if using any BAGLE parameterizations that implement Gaussian Processes (GPs), and \texttt{ephem} is required for loading ephemerides and other time and position data as needed.

\section{Microlensing Models in {\tt bagle.model}} \label{sec:models}

We consider  the case of a microlensing event between a point source and a point lens in
close proximity on the sky. 
The individual positions for the source and lens on the sky, as seen from Earth, are given by
\begin{eqnarray}
\Xsvecgeo(t) &= \Xsovec + \musvec[t - \tnot] + \pi_S \vect{P}(t, \alpha, \delta) \\
\Xlvecgeo(t) &= \Xlovec + \mulvec[t - \tnot] + \pi_L \vect{P}(t, \alpha, \delta) 
\end{eqnarray}
where $\Xsovec$ and $\Xlovec$ are the sky positions of the source and lens in the Solar System barycentric (SSB) coordinate system at time $t_0$ given in R.A. and Dec., $[\alpha, \delta]$.  Note, we use the Sun symbol, $\odot$, to indicate barycentric coordinates. The source and lens move with proper motion $\musvec$ and $\mulvec$ and the time $\tnot$ specifies the time of closest projected approach between the lens and source as seen from the SSB. Note that we adopt the standard convention that R.A.~increases to the East and Dec.~increases to the North. Finally, as seen from an Earth-centric view, their paths are modified by parallactic motion given by $\pi_S \vect{P}(t, \alpha, \delta)$ and $\pi_L \vect{P}(t, \alpha, \delta)$ where $\pi_S$ and $\pi_L$ are the maximum parallax amplitude given by $1/d_S$ and $1/d_L$, respectively, where $d_S$ is the distance to the source and $d_L$ is the distance to the lens (see \S\ref{sec:parallax} for more detail).
In the case of a fully photometric and astrometric BAGLE model, the intrinsic source and lens position are returned by Listing \ref{lst:lensAstrometry}.
\begin{lstlisting}[language=Python, caption={Python code for lens astrometry.}, label={lst:lensAstrometry}]
from bagle import model

par = <parameters, pseudo code>

# Instantiate BAGLE model.
pspl = model.PSPL_PhotAstrom_Par_Param1(par)

# Sample model at times, t.
t = np.arange(52000, 53000)

# Unlensed source position over time.
# Shape = [len(t), [E, N]] (e.g. len(t) x 2)
xS_unlens = pspl.get_source_astrometry_unlensed(t)

# Lens position over time.
xL = pspl.get_lens_astrometry(t)
\end{lstlisting}

\begin{figure}
    \centering
    \includegraphics[scale=0.5]{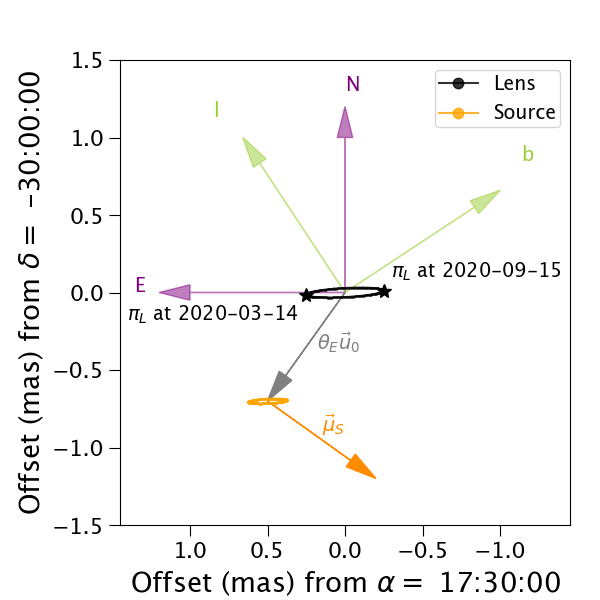}
    \caption{The basic geometry of a point-source, point-lens microlensing event, including parallax, toward the Galactic bulge as seen from Earth. The lens ({\em black}) and the unlensed source ({\em orange}) move in small parallax ellipses aligned with the ecliptic as seen from the geocentric perspective when at a distance of 4 kpc and 8 kpc, respectively.
    The compass rose for equatorial ({\em purple}) and galactic ({\em green}) coordinate systems are also shown.}
    \label{fig:lens_geometry}
\end{figure}

\begin{figure*}
    \centering
    \includegraphics[width=0.7\textwidth]{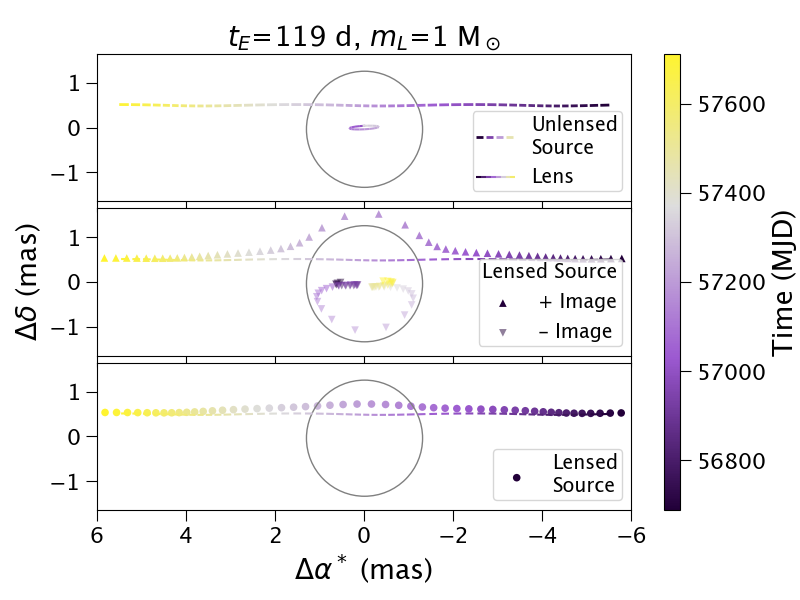}
    \caption{
    A BAGLE model for point-source, point-lens (PSPL) with parallax with 
    a dark lens of $M_L$ = 1 $\msun$, ~$\murel$ = 4 mas/yr, $\beta$ = 0.5 mas, $d_L$ = 3 kpc, $d_S$ = 8 kpc, and observed from Earth towards the bulge. 
    The {\em top} panel shows the position of the lens and the unlensed source. The {\em middle} panel shows the two sets of lensed source images. The {\em bottom} panel shows the flux-weighted centroid as would be observed on sky. The circle illustrates the Einstein radius ($\thetaE =$ 1.3 mas) centered on the lens position at time $t_0 =$ 57200 MJD. 
    \label{fig:pspl_ellipse}
    }
\end{figure*}

Microlensing primarily depends on the relative position of the source and lens with respect to each other. We shift into a relative coordinate system by defining
\begin{align}
\vect{X_\earth}(t) &= \Xsvecgeo(t) - \Xlvecgeo(t) \label{eq:rel_pos} \\
&= (\Xsovec - \Xlovec) + \murelvec\Delta t - \pirel\vect{P}(t) \nonumber
\end{align}
where $\Delta t = t - \tnot$, $\murelvec = \musvec - \mulvec$ and $\pirel \equiv -(\pi_S - \pi_L)$ is the relative parallax and is always a positive quantity. Figure \ref{fig:lens_geometry} shows the basic geometry, including parallax, of an example event where the source and lens both have no proper motion in order to illustrate the parallax ellipse orientations on the sky. 
In comparison, Figure \ref{fig:pspl_ellipse}, top panel shows the position of the source and lens where the source has a proper motion of 4 mas yr$^{-1}$.

% lens_math.plot_geometry()

% lens_math.plot_pspl_ellipse()

Introducing lensing into the picture, the characteristic scale for a microlensing event is the Einstein radius, in angular units on the sky (i.e. milli-arcseconds), given by

\begin{equation}
\thetaE = \sqrt{\frac{4 G M_L}{c^2}\left(\frac{1}{d_L} - \frac{1}{d_S}\right)}
\end{equation}
where $M_L$ is the mass of the lens and $d_L$ and $d_S$ are the distance
between the Sun and the lens and source, respectively. For simplicity, we will initially assume the lens is dark.

It is convenient to switch to a coordinate system normalized by the Einstein radius.
In any reference frame, we can redefine the relative positional offset as 
\begin{align}
\vect{u} &= \frac{\vect{X_S} - \vect{X_L}}{\thetaE}.
\end{align}
Note that this is the true angular offset and not the lensed position seen in the sky. The magnitude of the separation vector, $\vect{u}(t)$, between the source and the lens determines the amount of magnification and shift in direction as a function of time during the microlensing event.

When a lens and source are separated (in projection) by $\lesssim \thetaE$, the source will be gravitationally lensed, producing two lensed images indicated as $-$ and $+$ for the closer and further images in projection. 
The positions of the two lensed source images are
\begin{equation}
    \vect{X}_{\pm,S} = \left( u \pm \sqrt{u^2 + 4} \right) \frac{\thetaE}{2} \vect{\hat{u}}
\end{equation}
in angular units such as arcseconds. 
The offsets of the two images from the true (unlensed) source position are described by
\begin{equation}\label{eq:image_dc}
    \vect{\delta_{\pm}} = \left(-u \pm \sqrt{u^2 + 4} \right) \frac{\thetaE}{2} \vect{\hat{u}}.
\end{equation}
In a {\tt bagle} model, the positions of each lensed image on the sky as viewed from Earth are returned by Listing \ref{lst:res_ast}. Figure \ref{fig:pspl_ellipse}, middle panel shows the lensed image positions for an example microlensing system.
\begin{lstlisting}[language=Python, caption={Positions of lensed source images.}, label={lst:res_ast}]
# Shape = [len(t), [+,-], [E,N]]
xS_res = pspl.get_resolved_astrometry(t)
\end{lstlisting}
The amplification of each image is
\begin{equation}\label{eq:image_A}
    A_{\pm} = \frac{1}{2} \left( \frac{u^2 + 2}{u \sqrt{u^2 + 4}} \pm 1 \right).
\end{equation}
and is returned by Listing \ref{lst:res_amp}, again as seen from Earth.
\begin{lstlisting}[language=Python, caption={Amplification of lensed source images.}, label={lst:res_amp}]
# Shape = [len(t), [+,-]]
A_res = pspl.get_resolved_amplification(t)
\end{lstlisting}

In the case where the lensed images are unresolved (i.e. microlensing), the source will appear to increase in brightness, and its apparent position will be perturbed. 
We observe, from Earth, a total magnification of the combined images of
\begin{equation}
A(t) = \frac{u^2 + 2}{u \sqrt{u^2 + 4}},
\label{eq:mlens_A}
\end{equation}
which is returned by Listing \ref{lst:amp}.
\begin{lstlisting}[language=Python, caption={Total amplification.}, label={lst:amp}]
# Shape = [len(t)]
A_res = pspl.get_amplification(t)
\end{lstlisting}
The amount and direction of centroid offset, assuming a dark lens, is
\begin{equation}
\deltavec(t) = \frac{A_+\vect{\delta_+}+ A_-\vect{\delta_-}}{A_++A_-} = \frac{\vect{u}}{u^2 + 2} \thetaE
\label{eq:mlens_dc}
\end{equation}
from the source's true position on the plane of the sky.
{\tt bagle} returns the centroid shift due to lensing as seen from Earth using Listing \ref{lst:centroid_shift}. 
\begin{lstlisting}[language=Python, caption={Total astrometric shift due to lensing.}, label={lst:centroid_shift}]
# Shape = [len(t), [E,N]]
pos = pspl.get_centroid_shift(t)
\end{lstlisting}

The unresolved centroid position of the lensed source as observed on the sky from Earth, assuming a dark lens, is computed as the flux-weighted centroid of the two lensed source images in
\begin{equation}
\Xcentvec = \Xsvecgeo + \deltavec
\label{eq:mlens_Xcent}
\end{equation}
and is returned by {\tt bagle} in Listing \ref{lst:ast}. 
\begin{lstlisting}[language=Python, caption={Lensed astrometry seen on sky.}, label={lst:ast}]
# Shape = [len(t), [E,N]]
pos = pspl.get_astrometry(t)
\end{lstlisting}
The magnitude of the lensed source as observed on the sky from Earth, assuming a dark lens, is given by 
\begin{equation}
\magcent = -2.5 \log (F_S \cdot A) + ZP
\label{eq:mlens_mcent}
\end{equation}
where $F_S$ is the flux of the source and $ZP$ is the zeropoint to convert from flux to apparent magnitudes. 
{\tt bagle} returns the observed photometry as shown in Figure \ref{fig:pspl_phot_astrom}, top panel using Listing \ref{lst:phot}.
\begin{lstlisting}[language=Python, caption={Observed photometric light-curve.}, label={lst:phot}]
# Shape = [len(t)]
A_res = pspl.get_photometry(t)
\end{lstlisting}
Equations \ref{eq:mlens_A}, \ref{eq:mlens_dc},  \ref{eq:mlens_Xcent}, and \ref{eq:mlens_mcent} are only the observed values in the case of unresolved microlensing for a dark lens and no blended neighbor stars. 
However, the code Listings above works for luminous lenses as well (see \S\ref{sec:blending}). Figure \ref{fig:pspl_ellipse}, bottom panel, shows the astrometry as observed on sky for a dark lens.

\subsection{Microlens Event Characteristics}
\label{sec:event_characteristics}

The motions of the source and lens produce a transient photometric event with a characteristic timescale defined as the Einstein crossing time,
\begin{align} 
\tE = \frac{\thetaE}{\murel}
\label{eq:tE}
\end{align}
The transient photometric signal peaks at the ``closest approach'' distance, at time $\tnot$, as
\begin{equation}
\uveco = \frac{\Xsovec - \Xlovec}{\thetaE}.
\end{equation}
It is important to note that these equations are in the SSB reference frame and that the Einstein crossing time and closest approach time and distance change in  different frames (see \S\ref{sec:params_geo} for geocentric definitions).

There is also a transient astrometric signal; however, the peak photometric magnification and the peak astrometric offset \emph{do not} occur at the same time. 
Given Eq.~\ref{eq:mlens_dc}, the maximum astrometric shift of the unresolved centroid occurs when the source and lens are separated by

\begin{equation}
    u_{c,max,\odot} = \sqrt{2}
\end{equation}
and at time,
\begin{equation}
    t_{c,max,\odot} = \tnot \pm \tE \sqrt{2 - \uo^2}
\end{equation}
in the SSB frame. The amplitude of the maximum astrometric shift is
\begin{equation}
    \delta_{c,max,\odot} = \frac{u_{c,max,\odot} \theta_E}{u_{c,max,\odot}^2 + 2} = \frac{1}{\sqrt{8}}\theta_E.
\end{equation}

When fitting photometry-only observations of microlens events, it is convenient to re-parameterize Equation \ref{eq:rel_pos} in dimensionless units of $\thetaE$. 
Then, the amplitude of the relative proper motion in units of the Einstein radius is given by $1/\tE$ per Eq.~\ref{eq:tE}. 
To convert to an Earth-centric reference frame and to account for the Earth's acceleration around the SSB, we express the relative parallax due to Earth's motion in units of Einstein radii as
\begin{equation}
\piE = \frac{\pirel}{\thetaE}.
\end{equation}
The equation of relative motion as seen from Earth becomes
\begin{eqnarray}
\uvecgeo(t) &= \uveco + \frac{\Delta t}{\tE} \murelhat - \piE \vect{P}(t) \\
&= \uo \uvecohat + \tau \murelhat - \piE \vect{P}(t)
\end{eqnarray}
where $\tau = (t - \tnot)/\tE = \Delta t / \tE$.
Thus, even without knowledge of the absolute source and lens positions, proper motions, and parallaxes, both the amplifications and centroid offsets in units of $\theta_E$ can be determined from Equations \ref{eq:image_dc}-\ref{eq:mlens_dc}.
This is what is returned by all {\tt bagle} photometry-only models (e.g. {\tt PSPL\_Phot\_Par\_Param1}). 

The geometry of the lens--source--Sun (or SSB) system is set by  $\uvecohat$ and $\murelhat$, which are orthogonal unit vectors. They are typically expressed in the Equatorial coordinate system that increases to the East ($\vect{\hat{E}}$) and North ($\vect{\hat{N}}$) and  $\hat{z} = \vect{\hat{E}} \times \vect{\hat{N}}$ points away from the Solar System. 
A common convention in the photometric microlensing literature is to define the source-lens geometry with a new vector
\begin{equation}
    \piEvec = \piE \piEhat \;\; \textrm{where} \;\;
    \piEhat = \murelhat
\end{equation}
and to allow $\uo$ to have a sign such that
\begin{eqnarray}
\uo > 0 &\;\;\textrm{if}\;\; (\uvecohat \times \vect{\hat{N}}) \cdot \hat{z}  
& > 0 \\
\uo < 0 &\;\;\textrm{if}\;\; (\uvecohat \times \vect{\hat{N}}) \cdot \hat{z}
& < 0
\label{eq:u0_sign}
\end{eqnarray}
as first developed in \citet{Gould:2004}. In other words, when the source passes East (West) of the lens, $\uo$ is positive (negative). The unit-less closest approach distance, $\uo$, can also be expressed as an angular separation, $\beta = \uo \thetaE$.
These parameters and conventions are particularly important for modeling photometry-only data sets. 

\begin{figure}
    \includegraphics[width=0.5\textwidth]{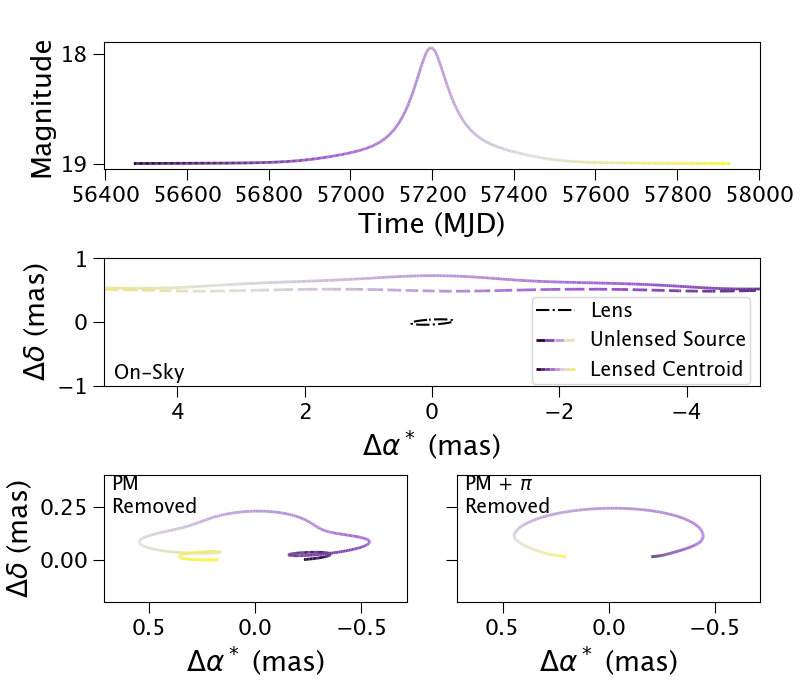}
    \caption{The on-sky photometry ({\em top}) and astrometry ({\em middle}) for the PSPL event from Figure \ref{fig:lens_geometry}.  The observed centroid position of the lensed source is shown as a solid line in the astrometric panels.
    Color indicates time as shown in the top panel horizontal axis. 
    The astrometry in different reference frames with the proper motion (PM) removed ({\em bottom left}) or PM and parallax removed ({\em bottom right}) show the astrometric microlensing signal traces an ellipse. 
    The complete set of parameters is
    $M_L$ = 1 M$_\odot$, $\beta$ = 0.5 mas, $X_{S,0}$ = [0.00, 0.50] mas, $\mu_S$ = [10, 0] mas/yr, $\mu_L$ = [0, 0] mas/yr, $d_L$ = 3 kpc, $d_S$ = 8 kpc, $b_{SFF}$ = 1, $mag_S$ = 19 mag, $t_0$ = 57200 MJD, $\alpha$ = 262.5 deg, $\delta$ = -30 deg.
    \label{fig:pspl_phot_astrom}
    }
\end{figure}

In the SSB lens-rest frame, the lensed source traces an ellipse during the lensing event with a major axis of
\begin{equation}
    a = \frac{\thetaE}{2\sqrt{\uo^2 + 2}}
\end{equation}
in the direction of $\murelhat$ (or $\piEhat$) and an ellipticity of
\begin{equation}
    e = \left( \frac{2}{\uo^2 + 2}\right)^{1/2}
\end{equation}
as shown in Figure \ref{fig:pspl_phot_astrom}, bottom-right panel.
In the geocentric frame, this ellipse is modulated by the parallax signal (Figure \ref{fig:pspl_phot_astrom}, left panel).

% lens_math.plot_pspl_photom_astrom()

\subsection{Parallax Vector Definition}
\label{sec:parallax}

Microlensing events with both photometric and astrometric signals provide much more information on the distances and parallaxes to the lens and source.
Thus, a careful treatment of parallax is essential. 
$\vect{P}(t, \alpha, \delta)$ is the direction and magnitude of the parallax
vector at a given time, $t$ and in the direction of the source-lens
system, ($\alpha$, $\delta$) given by the difference of the Earth and Sun's position, normalized by 1 AU. 
% Simplistically, $\vect{P}$ can be calculated with a coordinate projection given by
% \begin{align}
% P_{\alpha} =& \frac{R}{R_U}
% \left[ \cos \epsilon \cos \alpha \sin l - \sin \alpha \cos l \right]  \\
% P_{\delta} =& \frac{R}{R_U}
% [ (\sin \epsilon \cos \delta - \cos \epsilon \sin \alpha \sin \delta)
%              \sin l \\
% & - \cos \alpha \sin \delta \cos l ]
% \end{align}
% where $\epsilon$ is
% the obliquity of the ecliptic ($\epsilon = 23^\circ 27'$) and 
% $l$ is the ecliptic longitude of the
% sun on a specific date, and $R_U = 1$ AU \citep[\S3.4, Equation 3.20]{vanDeKamp:1967,Hog:1995}. 
% In the circular orbit approximation, the ecliptic
% longitude of the Sun is given by $l = 360^\circ (t + 284)/365$.
% %Perturbations from Jupiter can also be included by using routines such as the IDL \texttt{sun\_position} library.

Precise ephemerides are available in the Astropy package \citep{astropy:2013,astropy:2018,astropy:2022}. 
In BAGLE, we use the \texttt{get\_body\_barycentric\_posvel()} function in the \texttt{astropy.coordinates} library to compute $\vect{P}$.
First we compute the 3D position vector for the Sun and the Earth relative to the Solar System Barycenter, $\vect{Y_{\textrm{sun},SSB}}(t)$ and $\vect{Y_{\textrm{earth},SSB}}(t)$.
These 3D vectors are projected onto the tangential sky plane at the microlensing target position using,
\begin{eqnarray}
P_E =& [\vect{Y_{\textrm{sun},SSB}}(t) - \vect{Y_{\textrm{earth},SSB}}(t)] \cdot \vect{\hat{E}}_{3D} \\
P_N =& [\vect{Y_{\textrm{sun},SSB}}(t) - \vect{Y_{\textrm{earth},SSB}}(t)] \cdot \vect{\hat{N}}_{3D}
\end{eqnarray}
where $\vect{\hat{E}}_{3D}$ and $\vect{\hat{N}}_{3D}$ point East and North.

In BAGLE, we use JPL ephemerides \citep{Rhodes2011} through the Astropy interface, as they are more precise. 
The JPL ephemerides require a 115 MB file download on first use only.
Calls to calculate the parallax vector are also cached to increase computational efficiency for model fitting. The cache folder is, by default, limited to 1 GB and can be changed in \texttt{parallax.py}.
Alternatively, a user can set {\tt astropy} to use the built-in PyERFA orbital approximation \citep{marten_van_kerkwijk_2024_14072435} for the ephemerides instead as shown in Listing \ref{lst:eph_builtin}. 
\begin{lstlisting}[language=Python, caption={Python code to override the JPL ephemerides.}, label={lst:eph_builtin}]
from bagle import model

model.parallax.solar_system_ephemris.set('builtin')
\end{lstlisting}

BAGLE also provides the option for a non-Earth observer, via the JPL Horizons application\footnote{\url{https://ssd.jpl.nasa.gov/horizons/}} which provides ephemerides for a number of other Solar System objects and select spacecraft (including e.g.~Hubble, JWST, Gaia, and Euclid). 
To set a non-Earth observer, use the {\tt obsLocation} keyword in any {\tt bagle} model class at the time it is instantiated as shown in Listing \ref{lst:obsLocation}. 
\begin{lstlisting}[language=Python, caption={Python code to set a non-Earth observer.}, label={lst:obsLocation}]
par = <parameters, pseudo code>

# Instantiate BAGLE model.
pspl = model.PSPL_PhotAstrom_Par_Param1(par, obsLocation='jwst')
\end{lstlisting}

We have assumed that the orientation of the parallax vector is the same for both the source and the lens, which is a reasonable assumption for microlensing where the separations are small. 
We also ignore the small change in line-of-sight distance between the Earth/observer and source or lens throughout the year when converting from heliocentric to geocentric reference frames. 

\subsection{Blending}
\label{sec:blending}
An idealized microlensing event with an isolated source and a dark lens would magnify the source flux and shift the source position over time as described in Equations \ref{eq:mlens_A} and \ref{eq:mlens_dc}. 
However, in most microlensing events, the lens is luminous and dilutes the microlensing signal. 
Further, microlensing is most frequent in crowded fields, and the beam size or seeing of our telescopes typically includes multiple neighbor sources as well. 
To capture this contamination, we define ``blending" parameters such as the source flux fraction,
\begin{equation}
\bsff = \frac{F_S}{F_S + F_L + F_N}
\end{equation}
where $F_S$ is the unlensed flux of the source, $F_L$ is the lens flux, and $F_N$ is the flux from neighboring stars (including companions to the source or lens).
The photometric microlensing signal is then given as the total flux,
\begin{equation}
F_{cent} = A (t) F_S + \left( \frac{1 - \bsff}{\bsff} \right) F_S.
\end{equation}

The effect of blending is more complicated in the astrometric signal. 
The blended astrometric signal requires knowledge of not only the flux ratios but also the relative position and motion vectors of the source, lens and neighbors.
The astrometric shift vector for a microlensing event with a luminous lens and no luminous neighbors is
\begin{equation}\label{eq:astrom_blend}
         \vect{\delta}_{c, LL} = \frac{1 + g(u^2 + 3 - u\sqrt{u^2 + 4})}{(1 + g)(u^2 + 2 + gu \sqrt{u^2 + 4})} \theta_E \vect{u},
\end{equation}
where $g=F_L/F_S$ is the lens-source flux ratio (see Appendix \ref{sec:math_delta_ll}).

The observed positions on the sky for a microlensing event including a luminous lens is
\begin{equation}
\Xcentvec = \bsff \Xsvecgeo + (1 - \bsff) \Xlvecgeo + \vect{\delta}_{c, LL}.
\label{eq:ast_LL}
\end{equation}
The observed photometry on the sky, including a luminous lens, is
\begin{equation}
mag_{cent,\earth} = -2.5 \log \left(F_S \cdot A_{\earth} + F_S \frac{1 - \bsff}{\bsff}\right) + ZP.
\label{eq:phot_LL}
\end{equation}

At this time, BAGLE's astrometric models include blended light from a luminous companion in all model classes, and additional neighbors are not accounted for. However, all photometric models are able to handle blending from a luminous lens and/or other neighbors. Much higher spatial resolution is required for astrometric microlensing data sets than most photometric surveys provide, so blending from unrelated neighbors is reduced for the average astrometric data set in comparison to ground-based, non-adaptive optics (AO) photometric data. However, neighbor stars may still have an effect. Handling of luminous neighbor stars in astrometric models will be implemented in future BAGLE development as astrometric data sets improve and provide sufficient constraining power for the increased number of free parameters.

\subsection{Parameterization}
\label{sec:params}

In order to fully model a microlensing event, we must choose a means of parameterizing $\boldsymbol{u}$.
We typically choose to fit for heliocentric parameters for astrometric fits as the position at time $\tnot$ and proper motion of the source are naturally expressed in heliocentric coordinates (\S\ref{sec:params_helio}). However, most of the exoplanet microlensing community adopts a pseudo-geocentric reference frame as they predominantly work with short-timescale events and photometry only. We present conversions to this pseudo-geocentric frame in \S\ref{sec:params_geo}.

\subsubsection{Heliocentric}
\label{sec:params_helio}

In the heliocentric frame, the source-lens separation in units of $\thetaE$ is given by
\begin{align}
    \uvec(t) = \uveco + \left( \frac{t - \tnot}{\tE} \right) \murelhat
\end{align}
and the observed source-lens separation from Earth is
\begin{align}
    \uvecgeo(t) = \uvec(t) - \piE \vect{P}(t)    
\end{align}
    
The typical parameters used to fit a photometry-only microlensing event include:
\begin{itemize}
    \setlength\itemsep{-3pt}
    \item $\tnot$ (MJD)
    \item $\uo$ (including the sign)
    \item $\tE$ (days)
    \item $\piEvec = [\piEE, \; \piEN]_\odot = \piE \murelhat$
    \item $\bsff$
    \item $\magS = -2.5 \ln F_S + ZP$ (mag)
\end{itemize}
where the source flux fraction ($\bsff$) and source baseline magnitude ($\magS$) are defined for each filter and aperture, and a zero-point ($ZP$) is necessary for the conversion between magnitude and flux.
Some of the directional information in $\uvecohat$ is lost;
however, it can be recovered through the sign of $\uo$, which determines the parity between the direction of $\uveco$ and $\murelvec$ (Equation \ref{eq:u0_sign}).
In BAGLE, this parameterization is contained in \texttt{PSPL\_Phot\_Par\_Param1} and can be 
created as shown in Listing \ref{lst:param1}
\begin{lstlisting}[language=Python, caption={Python code for a photometric parameterization.}, label={lst:param1}]
from bagle import model

pspl = model.PSPL_Phot_Par_Param1(t0, u0, tE, 
                                  piEE, piEN, 
                                  bsff, magS, 
                                  raL=ra, 
                                  decL=dec)

# Plot photometry with daily samples
t_obs = np.arange(t0 - 10*tE, t0 + 10*tE, 1)
plt.figure()
plt.plot(t_obs, pspl.get_photometry(t_obs))
\end{lstlisting}

There are several alternative parameterizations and one particularly useful one is switching from $\magS$ to $mag_{base} = -2.5 \ln (F_S + F_L + F_N) + ZP$, since $mag_{base}$ and $\bsff$ are not degenerate with one another. 
$mag_{base}$ is the total baseline magnitude including both lens and source (and neighbor) flux.
In BAGLE, this parameterization is contained in \texttt{PSPL\_Phot\_Par\_Param2}.

Modeling joint photometry and astrometry data sets requires additional parameters. 
It is convenient to build off of the photometry-only parameters:
\begin{itemize}
    \setlength\itemsep{-3pt}
    \item $\tnot$ (MJD)
    \item $\uo$ (including the sign)
    \item $\tE$ (days)
    \item $\piEvec = [\piEE, \; \piEN]_\odot = \piE \murelhat$
    \item $\bsff$
    \item $\magS = -2.5 \ln F_S + ZP$ (mag)
    \item $\thetaE$ (mas)
    \item $\pi_S$ (mas)
    \item $\Xsovec = [X_{S,0,\odot,E}, X_{S,0,\odot,N}]$ (arcsec)
    \item $\musvec = [\mu_{S,\odot,E}, \mu_{S,\odot,N}]$ (mas yr$^{-1}$)
\end{itemize}
The additional astrometric parameters set the scale of the microlensing event with $\theta_E$ and describe the unlensed motion of the source.
In BAGLE, this parameterization is contained in \texttt{PSPL\_PhotAstrom\_Par\_Param2}.

Convenient alternative parameter sets include switching to the baseline magnitude instead of the source magnitude, $mag_S \rightarrow mag_{base}$, which can be found in \texttt{PSPL\_PhotAstrom\_Par\_Param4}. Further, it is sometimes convenient to fit in $\log_{10} \theta_E$, since the underlying mass function is a power-law. 
This can be found in \texttt{PSPL\_PhotAstrom\_Par\_Param3}. 
Lastly, for pure simulation work, it is often convenient to work in a purely physical parameter set that fully describes the source and lens distances and trajectories and the lens mass. In \texttt{PSPL\_PhotAstrom\_Par\_Param1}, the parameters are 
[$M_L$, $\tnot$, $\beta$, $d_L$, $d_L/d_S$, $\Xsovec$, $\mulvec$, $\musvec$, $\bsff$, $\magS$] where $M_L$ is the lens mass and $\beta$ is the closest approach distance in milli-arcseconds with the same sign convention as $\uo$ (i.e. $\beta = \uo \thetaE$).

Once a model is instantiated in BAGLE using the preferred parameterization, it is possible to call on Python functions to simulate photometric light curves and astrometric shifts. In our earlier Python example above (Listing \ref{lst:param1}), we demonstrated the set of commands necessary to plot photometry with daily samples. Now, we look at some of these function calls that can be made using BAGLE to retrieve the following properties of a microlensing event: (1) combined amplification of the sources, (2) astrometric shift, (3) centroid shift (Listing \ref{lst:plotting}).

\begin{lstlisting}[language=Python, caption={Python code for plotting BAGLE models.}, label={lst:plotting}]
from bagle import model

# Create model with photometric and astrometric parameters
# with parameters in heliocentric frame.
pspl = model.PSPL_PhotAstrom_Par_Param2(t0, u0, 
                    tE, thetaE, piS, 
                    piEE, piEN, 
                    xS0_E, xS0_N,
                    muS_E, muS_N,
                    bsff, magS, 
                    raL=ra, decL=dec,
                    obsLocation='earth')
                 
# Plot amplification each day as observed from Earth
t_obs = np.arange(t0 - 10*tE, t0 + 10*tE, 1)
plt.figure()
plt.plot(t_obs, pspl.get_amplification(t_obs))

# Plot the lens-induced astrometric shift
plt.figure()
shift = pspl.get_centroid_shift(t_obs)
plt.plot(t_obs, shift)

# Plot the astrometric shift amplitude
plt.figure()
shift_amp = np.linalg.norm(shift, axis=1)
plt.plot(t_obs, shift_amp)
\end{lstlisting}
\label{sec:bagle_calls_cont}

\begin{figure}
    \centering
    \includegraphics[width=1.0\linewidth]{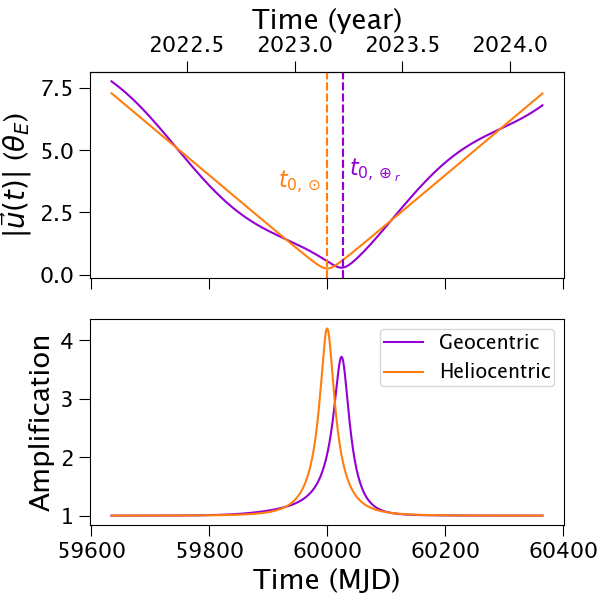}
    \caption{A microlensing event as viewed from heliocentric ({\em orange}) and true geocentric ({\em purple}) perspectives. We note that the heliocentric frame is rectilinear while the true geocentric frame is non-inertial. 
    The separation, $|\uvec(t)|$ ({\em top}), and amplification, $A(t)$ ({\em bottom}) illustrate the differences between $\uveco$ and $\uvecogeotr$ and $\tnot$ and $\tnotgeotr$.
    The complete list of heliocentric parameters for this event are $\alpha_L=262.5\degr$, $\delta_L=-30\degr$, $\tnot = 60000$, $M_L = 0.1 \msun$, $\beta = 0.1$ mas, $\Xsovec = [0.0, 0.1]$ mas, $\vect{\mu}_{L,\odot} = [0, 0]$ mas/yr, 
    $\vect{\mu}_{S,\odot} = [3, 0]$ mas/yr, 
    $d_L = 3$ kpc, $d_S = 8$ kpc, $\bsff = 1.0$.
    }
    \label{fig:to_amp_helio_vs_geo}
\end{figure}

Note that while the input parameters are in heliocentric coordinate systems, the output photometry and astrometry matches what would be observed from Earth, as indicated by the \texttt{obsLocation} parameter and shown in Figure \ref{fig:to_amp_helio_vs_geo}.

% lens_math.plot_helio_vs_geo_view()

\subsubsection{(Not quite) Geocentric: geo--$t_r$}
\label{sec:params_geo}

While the above heliocentric frame is the most general, it has become common convention in the microlensing literature to switch to a stationary, rectilinear frame that is geocentric at some reference time, $t_r$. 
In this frame, the parameters more closely mirror the properties of the observed photometric light-curve, such as the peak and width, at least for short-duration events with $\tEgeotr<50$ days (Figure \ref{fig:to_amp_helio_vs_geo}). 
Note that the convention in the microlensing literature is not a true geocentric reference frame and we will utilize $\geotr$ to indicate quantities in this new reference frame.
The reference time is commonly taken as the peak of the light curve or sometimes the $t_0$ derived from an initial non-parallax fit.

To transform into this geo-$t_r$ reference frame, a constant position and velocity offset are applied such that
\begin{align}
    \uvecgeotr(t) &= \uvec(t) - \piE \Pvec(t_r) 
    - (t - t_r) \piE  \Pdotvec(t_r) \\
     \murelvecgeotr &= \murelvec - \pirel \Pdotvec (t_r), \label{eq:murel_helio_geotr_conv}
\end{align}
where $\Pdotvec (t_r)$ is the instantaneous rate of change of the parallax vector at the reference time.
The observed source-lens separation in the geo-$t_r$ frame can then be expressed as
\begin{align}
    \uvecgeo(t) = & \; \uvecogeotr + \left( \frac{t - \tnotgeotr}{\tEgeotr} \right) \murelhatgeotr \nonumber \\
    &- 
    \piE \left[ \vect{P}(t) - \vect{P}(t_r) - (t - t_r) \Pdotvec (t_r) \right] 
    \label{ref:eq_eom_geotr}
\end{align}
and is derived fully in Appendix \S\ref{sec:math_geotr}. 
Conversions between the heliocentric and geo-$t_r$ frames are presented in Appendix \S\ref{sec:helio_geo_convert}
including \bagle~convenience functions and parameterizations in the geo-$t_r$ frame.
We note that \bagle's default $\tnot$ is in the heliocentric frame for most parameterizations and for internal storage, which is logical when incorporating astrometry into the models. 

\subsection{Finite Sources}
\label{sec:fspl}

The point-source, point-lens approximation is accurate for source stars with small angular radii of $\rho_*<<1$ where $\rho_* = \theta_{S,*} / \theta_E$ is the angular size of the source ($\theta_{S,*} = R_{S,*} / d_S$) in units of $\thetaE$ and $R_{S,*}$ is the physical radius of the star. 
For larger stellar radii, finite-source effects may become important, calling for the use of a finite-source point-lens (FSPL) model. BAGLE currently models a finite source as a uniform disk shape. In this case, we can avoid costly inverse ray-shooting grid calculations and instead follow the contour integration approach presented by \citet{Bozza2021}. Note that when $u_\odot(t) \leq \rho_*$, either the source itself or the arbitrary points on its source radius approach the lens caustic.  We found that this leads to skewed calculations of photometry and astrometry in regimes where $\uvec(t) \leq \rho_*$. To avoid erroneous results in such cases, we use an adaptive mesh grid in our contour integration calculations when $u_\odot \leq \rho_*$.

\begin{figure}
    \centering
    \includegraphics[width=1.0\linewidth]{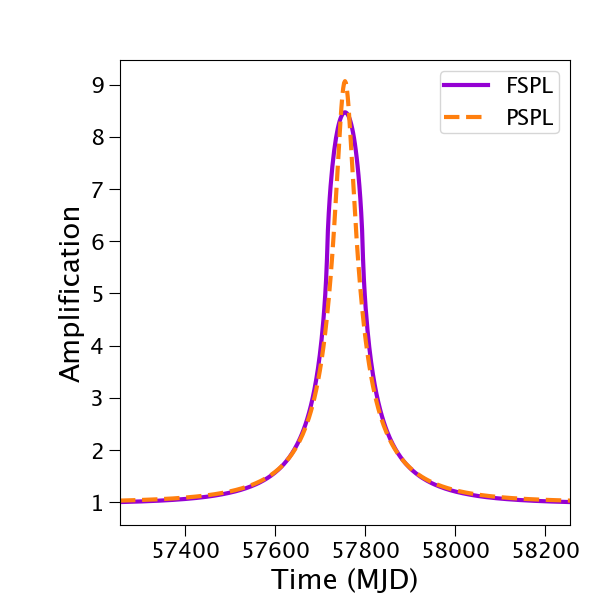}
    \caption{Comparison of finite-source, point-lens lightcurve with a point-source, point-lens lightcurve. The FSPL model was computed with n$_{outline}$=15 points along the source outline.
      The complete set of parameters is $M_L$ = 5 M$_\odot$, $\beta = 0.2$ mas, $\mu_L$ = [0, 0], $\mu_S$ = [4, 0], $\rho_*$ = 0.5 mas, $d_L$ = 4 kpc, $d_S$ = 8 kpc, $b_{SFF}$ = 1, $mag_S$ = 18 mag.
     }
    \label{fig:fs_photometry}
\end{figure}

A FSPL model is shown in Figure \ref{fig:fs_photometry} and can be created as shown in Listing \ref{lst:fspl}.
\begin{lstlisting}[language=Python, caption={Python code for a FSPL model.}, label={lst:fspl}]
fspl = model.FSPL_PhotAstrom_Par_Param1(mL, t0, 
            beta, dL, dL_dS, 
            xS0_E, xS0_N,
            muL_E, muL_N,
            muS_E, muS_N,
            radiusS, 
            b_sff, magS,
            raL = ra, decL = dec, 
            n_outline=50)
\end{lstlisting}

Details on the methods implemented for FSPL photometry (\S\ref{sec:FSPL_phot}) and astrometry (\S\ref{sec:FSPL_astrom}) are presented below.

\subsubsection{Finite Source Photometry}
\label{sec:FSPL_phot}

\begin{figure}
    \centering
    \includegraphics[width=1.0\linewidth]{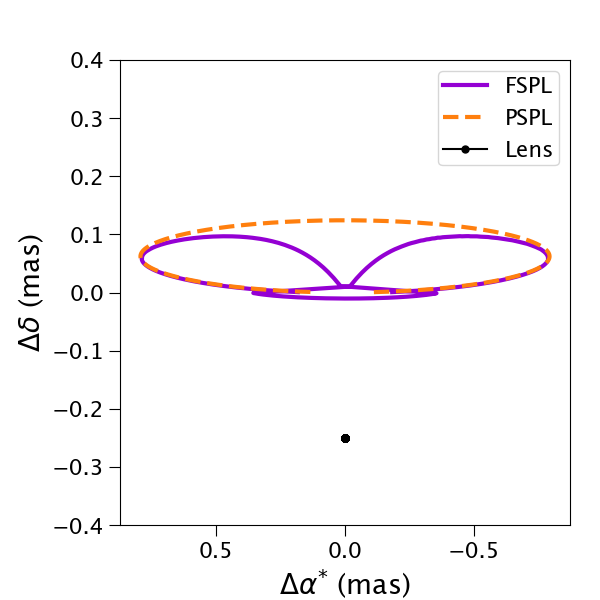}
    \caption{Comparison of the source's centroid shift between a finite-source and a point-source event.
      The same model parameters are used as in \autoref{fig:fs_photometry}. 
    }
    \label{fig:fs_astrometry}
\end{figure}

The true source position is represented by a circular outline of $n_{\rm outline}$ points, $\vect{y}$, corresponding to the positions at each angle $\phi$ around the star's circumference, evenly spaced by $\Delta\phi$. Surface brightness is preserved in microlensing; thus, a uniformly-lit disk model will produce the same uniform surface brightness in the warped, lensed images. 
Magnification can then be calculated as the total area of the lensed images divided by the area of the unlensed source.

The outlines of the major and minor image, $\vect{x_\pm}$, are computed by transforming each point in the true source outline contour to the source image outlines via the lens equation solutions given in Equation \ref{eq:image_dc}. We refer to each point in the contour as $x_{I,i,j}$, where $I$ indicates the image instance ($+$ or $-$), $i$ the point number (from 0 to $n_{\rm outline}$-1, inclusive), and $j$ indicates the directional component (E=East or N=North).

Following the procedure outlined by \citet{Bozza2021}, we use Green's theorem to convert the surface integral over the image to a line integral over the image boundary and compute the area of each image. To approximate this integral with a discrete set of image outline points via a trapezium approximation, we first follow their Equation (9), put in our terms as:

\begin{equation}
    \Omega_\pm = \mp \frac{1}{2} \sum_i (x_{\pm, i+1, N} + x_{\pm, i, N}) (x_{\pm, i+1, E} - x_{\pm, i, E}).
\end{equation}

Following their Equation (10), we add a parabolic correction that takes into account local curvature:

\begin{equation}
    d\Omega_\pm = \pm \frac{1}{24} \sum_i \left[ (\vect{x}'_\pm \wedge \vect{x}''_\pm)|_{\phi_i} + (\vect{x}'_\pm \wedge \vect{x}''_\pm)|_{\phi_{i+1}} \right] (\Delta\phi_i)^3,
\end{equation}
where the primes indicate the derivative of $\vect{x}$ in terms of $\phi$. This uses a wedge product, which for arbitrary 2-dimensional vectors $\vect{v}$ and $\vect{w}$ is,

\begin{equation}
\vect{v} \wedge \vect{w} = v_0 w_1 - v_1 w_0.
\end{equation}

From this, we can estimate the amplification of each image, 
\begin{equation}
    A_{\pm,\rm FS} = \frac{\Omega_\pm + d\Omega_\pm}{\Omega_*}, 
\end{equation}
where $\Omega_*=\pi \theta_*^2$ is the angular area of the star and $\theta_*$ is the source radius in mas.

The total amplification is simply $A_{\rm FS} = A_{+,\rm FS} + A_{-,\rm FS}$. We present a comparison between the finite-source photometry and point-source photometry in \autoref{fig:fs_photometry}.

\subsubsection{Finite Source Astrometry}
\label{sec:FSPL_astrom}

We calculate the astrometric positions of each lensed image centroid via the \citet{Bozza2021} Equations (19-20), again using the Green's theorem approximation via the discrete image outlines. 
For each lensed image and dimension, we define
\begin{align}
Y_{\pm, E} &= \pm \frac{1}{8}\sum_i (x_{\pm,i+1,E} + 
    x_{\pm,i,E})^2 (x_{\pm, i+1,N} - x_{\pm,i,N}) 
    \\
Y_{\pm, N} &= \mp \frac{1}{8}\sum_i (x_{\pm,i+1,N} + 
    x_{\pm,i,N})^2 (x_{\pm,i+1,E} - x_{\pm,i,E}).
\end{align}
Following their Equation (21) and (22), we add a parabolic correction to the above astrometric terms,
\begin{align}
        dY_{\pm, E} &= \pm \frac{1}{24}\sum_i \bigg[ \bigg(x'^{2}_{\pm,E} x'_{\pm,N} + x_{\pm,E}(\vect{x'}_{\pm}\wedge \vect{x''}_{\pm})\bigg)|_{\phi_i} \nonumber \\ 
        &+\bigg(x'^{2}_{\pm, E} x'^{}_{\pm, N} + x_{\pm, E}(\vect{x'}_{\pm}\wedge \vect{x''}_{\pm})\bigg)|_{\phi_{i+1} }\bigg] (\Delta\phi_i)^3 
        \\
        dY_{\pm, N} &= \pm \frac{1}{24}\sum_i \bigg[ \bigg(x'^{2}_{\pm, N} x'^{}_{\pm, E} + x_{\pm, N}(\vect{x'}_{\pm}\wedge \vect{x''}_{\pm})\bigg)|_{\phi_i}
        \nonumber \\  &+\bigg(x'^{2}_{\pm, N} x'^{}_{\pm, E} + x_{\pm, N}(\vect{x'}_{\pm}\wedge \vect{x''}_{\pm})\bigg)|_{\phi_{i+1} }\bigg] (\Delta\phi_i)^3.
\end{align}
The centroid for each lensed image is then
\begin{equation}
    \vect{X_{\pm,FS}} = \frac{\vect{Y_{\pm}} + \vect{dY_{\pm}}}
                        {\Omega_\pm + d\Omega_\pm}
\end{equation}

Overall, the flux-weighted image centroid ($\vect{X_{cent,FS, lensed}}$) for finite sources can be found using $\vect{X_{\pm, FS}}$, $A_{\pm, FS}$, the flux of the source ($F_S$), the flux of the lens($F_L$) and the position of the lens $\Xlvec$ as follows:
\begin{align}
   \vect{X_{cent,FS, lensed}} =
   \frac{ F_L \Xlvec \; + \displaystyle \sum_{I=\{+, -\}} F_S A_{I, FS} \vect{X_{I, FS}}}
   {F_L \; + \displaystyle \sum_{I=\{+, -\}} F_S A_{I, FS}}
\end{align}

%Overall, the flux-weighted image centroid ($\vect{X_{cent,FS, lensed}}$) can be found by integrating the point-source solution over the finite source disk, presented in Equation \ref{eqn:fs_centroid_solution} as: 

%\begin{align}
   % \label{eqn:fs_centroid_solution}
   %\vect{X_{cent,FS, lensed}} =
   %\frac{F_S \displaystyle \int  \bigg(\displaystyle \sum_{i=\{+, -\}} A_{i, }\vect{X_{i}}\bigg) dA_{source}+F_L \Xlvec}{\displaystyle F_S \int \bigg(\displaystyle \sum_{i=\{+, -\}} A_{i}\bigg ) dA_{source} + F_L}
%\end{align}

%where $\vect{X_{\pm}}$ is the positions of the two lensed point-source images,$A_{\pm}$ are the point-source amplifications for each image, $F_S$ is the flux of the source, $F_L$ is the flux of the lens and $\Xlvec$ is the position of the lens. 

We present a comparison between the centroid shift of a finite-source and point-source event in \autoref{fig:fs_astrometry}.
 
 Future work on BAGLE will include the addition of a limb-darkened FSPL model and a binary-lens FSBL model.

\section{BAGLE Model Implementation}
\label{sec:bagle_implemenation}

BAGLE provides a set of classes and functions that allow the user to construct microlensing models. The available classes for instantiating a microlensing event model are shown
in the API documentation\footnote{\url{https://bagle.readthedocs.io/en/latest/}}.

Note, each model class has a name that typically has a structure of:
\begin{lstlisting}[language=Python]
<ModelDataType>_<Parallax>_<GP>_<Parameterization>.
\end{lstlisting}
For example, {\tt PSPL\_Phot\_noPar\_Param2} has a data and model class type of {\tt PSPL\_Phot},
which contains a point-source, point-lens event with photometry only. The model
has no parallax, no gaussian process (GP) correlated noise and uses parameterization \#2.

Each model class is built up from a menu of different features
by inheriting from multiple base classes, each from a different {\em family} of
related classes (Figure \ref{fig:bagle}).
Each microlensing Model Class must contain a Data Class (e.g. {\tt PSPL\_Phot} for a photometry-only point-source, point-lens model), a Parallax Class (e.g. {\tt PSPL\_noParallax} to ignore parallax), and a Parametrization Class (e.g. {\tt PSPL\_PhotParam1}). Optionally, it may also include a GP (Gaussian process) Class (e.g. {\tt PSPL\_GP}).

For example, the {\tt PSPL\_PhotAstrom\_noPar\_Param1} model is declared as:
\begin{lstlisting}[language=Python]
class PSPL_PhotAstrom_noPar_Param1(ModelClassABC, PSPL_PhotAstrom, PSPL_noParallax, PSPL_PhotAstromParam1)
\end{lstlisting}
Each class family is described in more detail below.

\subsection{Model Class Family}
A Model Class can be instantiated by a user, effectively working as a container to bring together the Data, Parallax, Parametrization, and optional GP Classes.
The base Model Class is ModelClassABC.

\subsection{Data Class Family}

The Data Classes inform the model of what type of data will be used.
If the model will be for photometry only, then a model with the {\tt PSPL\_Phot}
class must be selected. These models have the words {\tt Phot} in their names.
If the model will be using photometry and astrometry data, then a model with
the {\tt PSPL\_PhotAstrom} must be selected. These models have the words
{\tt PhotAstrom} in their names. Likewise, the {\tt PSPL\_Astrom} classes are for astrometry-only data, and their names include {\tt Astrom}.

Data containing astrometry will generate a warning that astrometry data will
not be used in the model when run through a model using `PSPL\_Phot' and vice versa. Data that
does not contain astrometry run through a model using `PSPL\_PhotAstrom' will
generate a RuntimeError.

The base Data Class is {\tt PSPL}.

\subsection{Parallax Class Family}

The Parallax classes set whether the model uses parallax when calculating
photometry, calculating astrometry, and fitting data. There are only two
options for this class family, {\tt PSPL\_noParallax} and {\tt PSPL\_Parallax}. Models
that do not have parallax have the words {\tt noPar} in their names, while models
that do contain parallax have the words {\tt Par} in their names.

The base Parallax Class is {\tt ParallaxClassABC}.

\subsection{Parameterization Class Family}

The Parametrization Classes determine which physical parameters define the model. 
Each parameterization class should have an {\tt \_\_init\_\_()} function defined that accepts input parameters and converts them into standard set of internal variables used throughout calculations. 
These internal variables are available for the user to access once a model class is instantiated (Table \ref{tab:bagle_pspl_variables}). 
The base Parametrization Class is {\tt PSPL\_Param}.

\begin{deluxetable*}{llll}
\tabletypesize{\footnotesize}
\tablewidth{0pt}
\tablecaption{BAGLE PSPL Class Variables \label{tab:bagle_pspl_variables} }
\tablehead{
\colhead{Name} & 
\colhead{Quantity} &
\colhead{Units} & 
\colhead{Notes}
}
\startdata
\texttt{raL} & $\alpha_L$ & deg & Right ascension of the microlensing event. \\
\texttt{decL} & $\delta_L$ & deg & Declination of the microlensing event. \\
\texttt{mL} & $M_L$ & $\msun$ & Lens mass. \\
\texttt{thetaE\_amp} & $\thetaE$ & mas & Einstein radius. \\
\texttt{t0} & $\tnot$ & MJD & Time of closest approach in heliocentric reference frame. \\
\texttt{beta} & $\beta$ & mas & Offset between the source and lens at time $\tnot$ in the heliocentric frame. \\
\texttt{u0\_amp} & $\uo$ & $\thetaE$ & Offset between the source and lens at time $\tnot$, normalized by the Einstein radius. \\
\texttt{xS0} & $\Xsovec$ & arcsec & Position of source on sky in [E, N] relative to the reference [$\alpha_L$, $\delta_L$] at time $\tnot$. \\
\texttt{xL0} & $\Xlovec$ & arcsec & Position of lens on sky in [E, N] relative to the reference [$\alpha_L$, $\delta_L$] at time $\tnot$. \\
\texttt{thetaS0} & $\Xsovec - \Xlovec$ & arcsec & Vector offset between the source and lens at time $\tnot$ in the heliocentric frame. \\
\texttt{muL} & $\mulvec$ & mas yr$^{-1}$ & Proper motion vector of the lens in the heliocentric frame. \\
\texttt{muS} & $\musvec$ & mas yr$^{-1}$ & Proper motion vector of the source in the heliocentric frame. \\
\texttt{muRel} & $\musvec - \mulvec$ & mas yr$^{-1}$ & Relative S-L proper motion vector. \\
\texttt{muRel\_amp} & $|\musvec - \mulvec|$ & mas yr$^{-1}$ & Relative S-L proper motion amplitude. \\
\texttt{muRel\_hat} & $\hat{\murelvec}$ &  & Normalized vector of the direction of the relative proper motion. \\
\texttt{u0\_hat} & $\uvecohat$ &  & Normalized vector of the direction of $\uo$ at time $\tnot$ in the heliocentric frame. \\
\texttt{dL} & $d_L$ & pc & Distance to the lens. \\
\texttt{dS} & $d_S$ & pc & Distance to the source. \\
\texttt{piL} & $\pi_L$ & mas & Parallax of the lens. \\
\texttt{piS} & $\pi_S$ & mas & Parallax of the source. \\
\texttt{piRel} & $\pirel$ & mas & Relative parallax between the lens and source. \\
\texttt{piE\_amp} & $\piE$ & & Microlensing parallax. \\
\enddata
\end{deluxetable*}

\subsection{GP Class Family (optional)}

The GP Classes optionally add in Gaussian processes to model correlated noise in microlensing data for one or more photometric data sets. \citet{Chen:2025} will present the \bagle~GP implementation in full detail.

\section{BAGLE Model Fitting}
\label{sec:bagle_fits}

BAGLE can simultaneously fit photometric and astrometric data using the PyMultiNest package \citep{pymultinest}, which implements a nested sampling fitting technique. 
Many parameters in microlensing introduce degeneracies and multi-modal solutions into the parameter space such as the degeneracy between $\pm u_0$.
We use nested sampling to fully explore degeneracies, find and separate multi-modal solutions, and fully sample the posterior probability distributions. 
Nested sampling also has the advantage of returning the Bayesian evidence, which allows for hypothesis testing and comparisons across different model classes (e.g. with and without parallax). 

In order to fit a microlensing event using BAGLE we instantiate a model class object to fit and one or more data set(s), where the data sets must be loaded into a data dictionary of type {\tt bagle.data.EventDatadict}.
As an example, a fitter object for test data set to be fit to the \texttt{PSPL\_PhotAstrom\_Par\_Param2} model can be set up via:

\begin{lstlisting}[language=Python]
from bagle import model, model_fitter, fake_data

data, _ = fake_data.fake_data1()
fitter = model_fitter.MicrolensSolver(data, model.PSPL_PhotAstrom_Par_Param2)

\end{lstlisting}
Properties of the data and fitter are described in more detail below. 

\subsection{Data Dictionary}
\label{sec:data_dict}

The {\tt EventDataDict} object is a python dictionary that can hold multiple data sets for a single microlensing event. 
The object must be initialized with
\begin{itemize}
    \setlength\itemsep{-3pt}
\item {\tt target}: event name
\item {\tt raL}: R.A. of the target
\item {\tt decL}: Dec. of the target.
\end{itemize}
For each photometric data set, the dictionary should contain:
\begin{itemize}
    \setlength\itemsep{-3pt}
\item {\tt t\_phot1}: Numpy array of times in MJD.
\item {\tt mag1}: Numpy array of magnitudes.
\item {\tt mag\_err1}: Numpy array of magnitude errors. 
\item {\tt t\_phot2}:  ... additional data sets...
\item {\tt mag2}:     ... additional data sets...
\item {\tt mag\_err2}: ... additional data sets...
\end{itemize}
where the `1' at the end is incremented for additional photometric data sets. 

For each astrometric data set, the dictionary contains:
\begin{itemize}        
    \setlength\itemsep{-3pt}
\item {\tt t\_ast1}: Numpy array of times in MJD.
\item {\tt xpos1}: Numpy array of positions on the sky in the East direction in arcsec.
\item {\tt ypos1}: Numpy array of positions on the sky in the North direction in arcsec.
\item {\tt xpos\_err1}: Numpy array of positional errors on the sky in the East direction in arcsec.
\item {\tt ypos\_err1}: Numpy array of positional errors on the sky in the North direction in arcsec.
\item {\tt t\_ast2}:    ... additional data sets...
\item {\tt xpos2}:     ... additional data sets...
\item {\tt ypos2}:     ... additional data sets...
\item {\tt xpos\_err2}: ... additional data sets...
\item {\tt ypos\_err2}: ... additional data sets...
\end{itemize}
where the index is incremented for additional astrometric data sets.

There should also be entries containing a list of the names and file locations of all the photometric and astrometric data sets loaded. 
This is used during fitting and for reporting (and convenient reloading).
The entries are:
\begin{itemize}
    \setlength\itemsep{-3pt}
\item {\tt phot\_data}: python list of names for photometric data sets (e.g.~`I\_OGLE', `Kp\_Keck', `Ch1\_Spitzer', `MOA')
\item {\tt phot\_files}: list of filename strings for photometric data sets.
\item {\tt ast\_data}: list of names for astrometric data sets (e.g. `Kp\_Keck')
\item {\tt ast\_files}: list of filename strings for astrometric data sets.
\end{itemize}
These entries {\em must} match the order and length of the corresponding data sets. 

Note, if a data set is used to provide both photometry and astrometry, it should have the same {\tt phot\_data} and {\tt ast\_data} name. This will be used to tie parameters together between the two sets during fitting. 
The {\tt bagle.data} module also contains convenience functions for reading in common file types from OGLE, MOA, and KMTNet. 

\subsection{Priors for Fitting}
\label{sec:priors}
For effective nested sampling, we must set appropriate priors. The full list of prior generators is supplied in the API documentation, including a number of generic functions and some specific functions for certain parameters.
As an example, a uniform prior on $t_0$ can be set via:

\begin{lstlisting}[language=Python,upquote=True]
fitter.priors['t0'] = model_fitter.make_gen(55750, 55800)
\end{lstlisting}
BAGLE includes default priors for every parameter, but we recommend manually setting one's own to explore all and only reasonable parameter space on an individual basis by event. This is particularly important for complex model fits and large data sets, where the fitter process becomes computationally expensive.

BAGLE includes the capability to set uniform priors (\texttt{make\_gen}), normally distributed priors (\texttt{make\_normal\_gen}), log-normally distributed priors (\texttt{make\_lognormal\_gen} and \texttt{make\_lognormal10\_gen}), truncated lognormally distributed priors, a number of priors for specific parameters based on the data (i.e. baseline magnitude based on a sigma clipped mean of the dataset), and some based on distributions from simulated microlensing surveys. Custom priors can also easily be created. All the priors can be found in the \texttt{model\_fitter} module. 

\subsection{Running a Model Fit}
\label{sec:fun_fit}

The nested sampling fit can be run via 
\begin{lstlisting}[language=Python]
fitter.solve() 
\end{lstlisting}
This will automatically generate several plots, including residuals, traces, and corner plots as the fit proceeds.
Most \texttt{pymultinest} parameters are exposed on instantiation of the fitter object (e.g. \texttt{n\_live\_points}).

\bagle~ provides several functions to load and examine the fit results. As an example, the following line will load the maximum likelihood best fit as a Model object:
\begin{lstlisting}[language=Python, upquote=True]
# Get the best fit
model_fit = fitter.get_best_fit_model(def_best='maxl')
\end{lstlisting}
The best-fit model parameters can be returned based on the maximum likelihood solution, the maximum aposteriori solution, or the mean or median of the marginalized 1D posterior probability distribution for each parameter. 
In the mean case, parameter uncertainties are estimated as the standard deviation of the marginalized 1D posterior.
In the median case, 68.3\%, 95.5\% and 99.7\% credible intervals are calculated from the marginalized 1D posterior.

The \bagle~ model fitter includes three schemes for weighting different data sets. The default {\tt MicrolensSolver} gives equal weight to each {\it point} in all the data sets. In some cases, e.g.~when including astrometric data, which tends to have a significantly smaller data set, important data can be outweighed by the large data sets from ground-based microlensing surveys. In such a case, it may be optimal to instead use the {\tt MicrolensSolverWeighted}, which has three options: `phot\_ast\_equal', `all\_equal', or custom user-input weights. In the default case, `phot\_ast\_equal', the total weighting of all photometric data sets is equal to the total weighting of all astrometric data sets. For `all\_equal', each data set is weighted equally, such that, e.g., one astrometric data set would be weighted equally to one photometric data set, regardless of the number of data points in each set or data sets per category. A user may alternatively input an array with a custom weight for each data set. 
%Finally, the {\tt MicrolensSolverHobsonWeighted} class implements a hyperparameter that controls the weighting of each data set, as described by \citet{Hobson2002}. 
{\tt MicrolensSolver} and the alternatively weighted versions can be used to fit any BAGLE model.

Runtimes for the \bagle~ model fitter are highly dependent on the data volume, data set number, and model complexity. For example, a PSPL model with parallax, one photometric data set, and reasonable but not extremely strict priors can run single-threaded in roughly a couple of minutes. Binary source, binary lens models with several photometric and astrometric data sets run multi-threaded can take days or even weeks to complete. 
Example run-times in some simple cases are presented in \S\ref{sec:results_fit_runtimes}. 

Also note that the \texttt{MicrolensSolver} object has convenient likelihood and prior probability functions that can be based into alternate optimization engines such as \texttt{dynesty} \citep{dynesty} and \texttt{emcee} \citep{emcee}, although we have found \texttt{pymultinest} to be the most robust and accurate.
Planned future work on \bagle~includes implementing a faster gradient-based fitter.

\section{Validation of Models}
\label{sec:validation}

In this section, we compare BAGLE with other microlensing models presented in the literature and with other packages, i.e., VBMicrolensing, pyLIMA, and MulensModel. 
The comparison is limited to point-source point-lens events, reserving a comparison between binary lenses and sources for future work \citep{Bhadra:2025}. 

\begin{figure}[b]
    \centering
    \includegraphics[width=0.43 \textwidth]{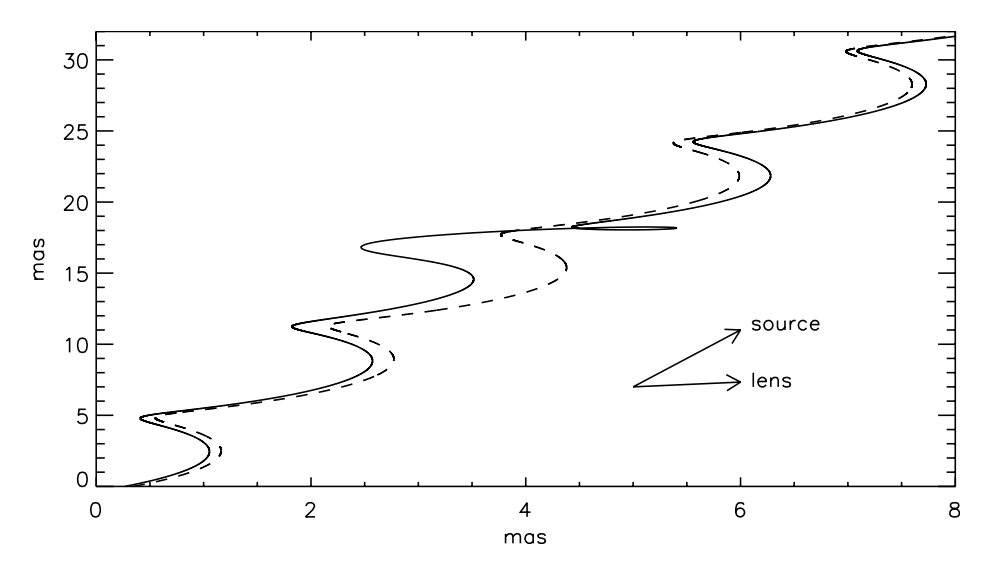}
    \hspace*{-0.38in} \includegraphics[width=0.49\textwidth]{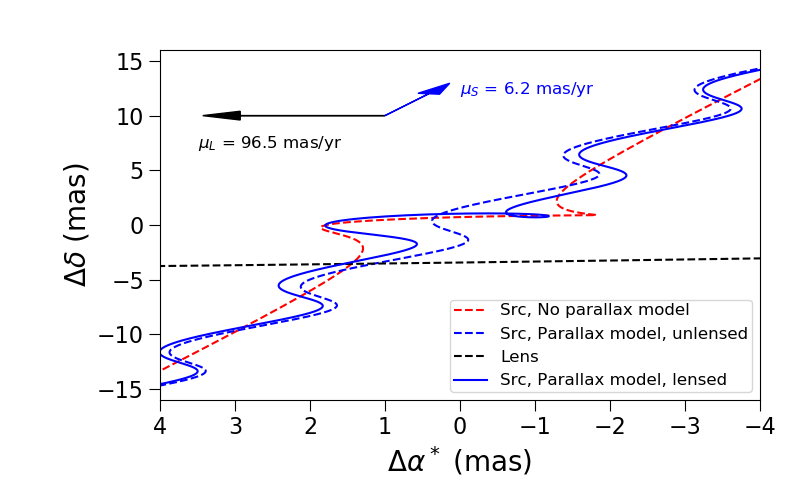}
    \caption{
    {\em Left:} Reproduced from Figure 1 of \citet{Belokurov:2002}. The relative 
    right ascension ($\alpha$) and declination ($\delta$) of a source are
    shown both with ({\em solid}) and without ({\em dashed}) a
    microlensing event with detectable parallactic contributions. 
    {\em Right:} New PSPL model with parallax reproduces previous results both with 9{\em blue solid}) and without ({\em blue dashed}) a lensing event, after accounting for a reference frame flip along the East direction.
    }
    \label{fig:test_belokurov_pub}
\end{figure}

First, we compare to an example PSPL event published in \citet{Belokurov:2002} as shown in Figure \ref{fig:test_belokurov_pub}.
The published parameters for this test are:
\begin{itemize}
\setlength\itemsep{0em}
\item $M_L = 0.5$ M$_\odot$
\item $d_L = 150$ pc
\item $d_S = 1500$ pc
\item $\tilde{v} = 70$ km/s (transverse velocity)
\item $\beta$ = -1.5 mas ($u_0$ = 0.303)
%\item $u_0 = -7.41$ mas
\end{itemize}

Note that $[\alpha, \delta]$ are not specified, nor are the direction
of the proper motion vectors, $\vec{\mu}_S$ and $\vec{\mu}_L$, given. We assume
a bulge microlensing event with $\alpha = 17.5$ h and $\delta =
-30.0^\circ$. The proper motions can be partially inferred from the
arrows on Figure \ref{fig:test_belokurov_pub}, and we adopt
$\vec{\mu}_S = [-1.75, 6.0]$ mas yr$^{-1}$ and $\vec{\mu}_L = [96.51, 0.0]$ mas yr$^{-1}$ based on these arrows and the transverse velocity. 
We assume no blending ($\bsff = 1.0$), a source baseline magnitude of $I=19$ mag, and a time of closest approach of $\tnot = 57290$ MJD. 
In order to reproduce these results, the following modifications were made. 
\citet{Belokurov:2002} listed the simulation as $u_0 = 1.5$; however, we believe this was actually $\beta = -1.5$ mas and $u_0$ = 0.303. 
Further, the lens proper motion was modified to point in the opposite East-West orientation as the source proper motion.
As shown in Figure \ref{fig:test_belokurov_pub}, \bagle~produces a similar trajectory on the sky for both the unlensed and lensed source position.

Now we compare to other contemporary microlensing modeling packages. First, a BAGLE model was created using the parameterization \texttt{PSPL\_PhotAstrom\_noPar\_Param2} and the following heliocentric event parameters: $\tnot = 55775$ MJD, $\tE = 100$ days, and $\uo = 1.0$. 
In the comparison of events with parallax, two additional input parameters are provided, i.e., the microlensing parallax in the East and North coordinates as $\piEvec = [0.2, 0.1]$. The event is placed at $\alpha_L=269.271\degr$ and $\delta_L-30.383\degr$. 
Since most packages other than \bagle~ prefer geo-$t_r$ parameters (explored further in \S\ref{sec:params_geo}), it is necessary to convert our heliocentric quantities to the geo-$t_r$ frame of reference (see Appendix \ref{sec:helio_geo_convert}). 
The resulting geo-$t_r$ parameters are $\tnotgeotr = 55773.483$ MJD, $\tEgeotr = 41.474$ days, $\uogeotr = 0.914$, and $\piEvecgeotr = [-1.242, -0.280]$. When calculating photometry with no parallax, this conversion is not necessary.

\begin{figure}
    \centering
    \includegraphics[width=0.48 \textwidth]{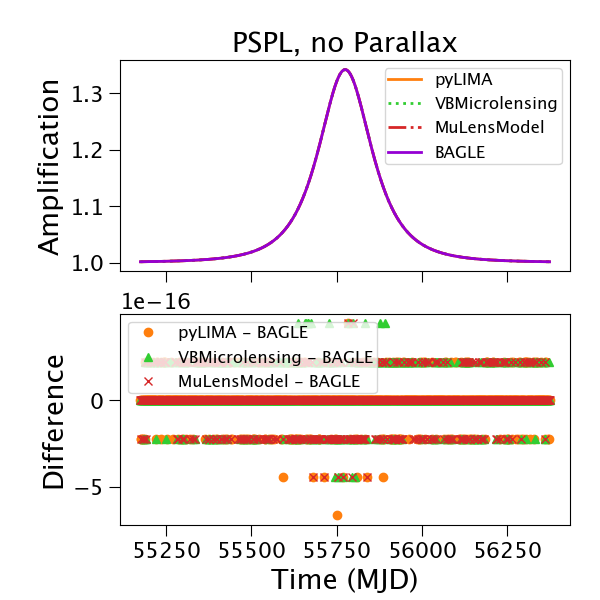}
    \caption{Comparison of a simulated PSPL event without parallax between VBMicrolensing, pyLIMA, MuLensModel, and BAGLE. The amplification ({\em top}) and residuals with respect to the BAGLE model ({\em bottom}) are shown over time. The differences in lightcurves are at machine precision. The heliocentric model parameters for this event are
    $\tnot = 55775$ MJD, $\uo = 1$, $\tE = 100$ days, $\bsff = 1$, $\magS = 19$ mag.
    }
    \label{fig:comparison_nopar}
\end{figure}

\begin{figure}
    \centering
    \includegraphics[width=0.48 \textwidth]{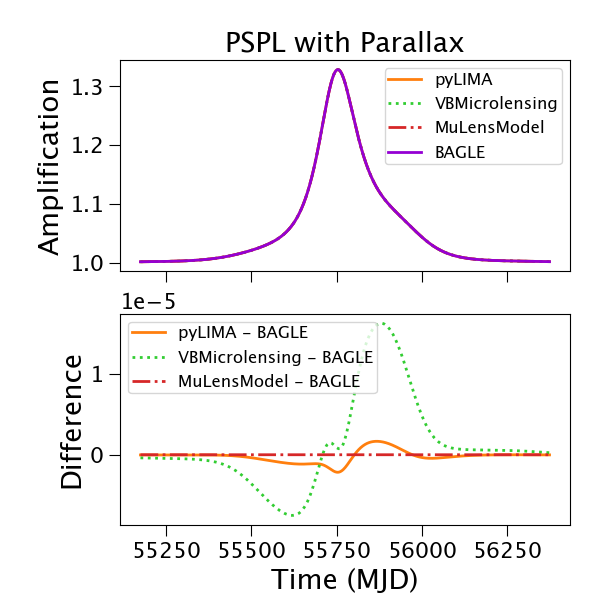}
    \caption{Comparison of a simulated PSPL event's amplification between VBMicrolensing, pyLIMA, MuLensModel, and BAGLE with parallax. The comparison was done over $t_{E,\odot}=100$ days with $\piEE = 0.2$ and $\piEN = 0.1$.}
    \label{fig:comparison_par}
\end{figure}

The comparison results are presented for events without parallax in \autoref{fig:comparison_nopar} and for events with parallax in \autoref{fig:comparison_par}.
In \autoref{fig:comparison_nopar}, the comparison between simulated light-curves from BAGLE, MuLensModel, and pyLIMA demonstrates minor residual differences at an order of magnitude of $10^{-16}$, which is consistent with our numerical precision. 
However, when including parallax, we observe more significant differences (\autoref{fig:comparison_par}) in residuals at the 10$^{-5}$ level, likely due to differences in the parallax implementations. 

Next, we compare BAGLE's astrometry with the astrometry calculations produced by other microlensing packages. This comparison is limited to VBMicrolensing and pyLIMA since these two packages are capable of reproducing centroid trajectories for sources and lenses. 
When comparing astrometry, there is an additional parameter for the source's proper motion: $\musvec = [0.2, 0.1] \frac{mas}{yr}$. The source's proper motion is input in heliocentric quantities by all three packages. 

\begin{figure}
    \centering
    \includegraphics[width=0.48 \textwidth]{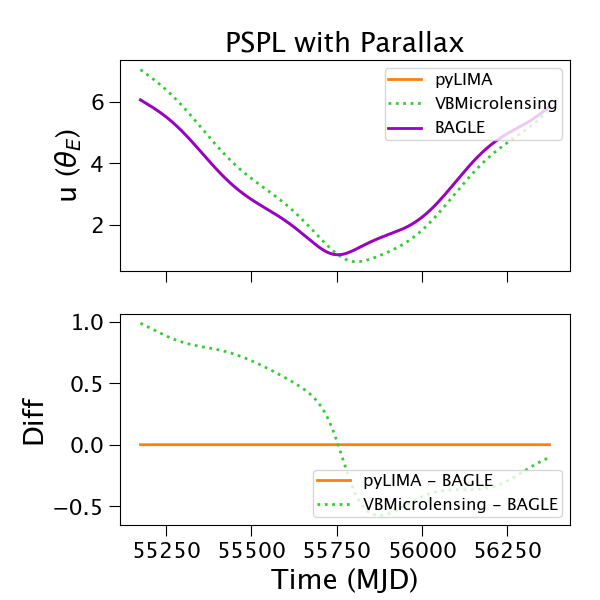}
    
    \caption{Comparison between BAGLE, VBMicrolensing, and pyLIMA for a simulated PSPL+parallax event's normalized separation, $u(t)$.
    The normalized source-lens separation over time is shown ({\em top}) along with the residuals ({\em bottom}).
    We use the same microlensing parameters as in Figure \ref{fig:comparison_par}: $\tnot = 55775$ MJD, $\uo = 1$, $\tE = 100$ days, $\piEE=0.2$, $\piEN=0.1$, $\bsff = 1$, $\magS = 19$ mag.}
    \label{fig:comparison_phot_astrom_par_umag}
\end{figure}

We compare the normalized source-lens separation as a function of time for the three packages in \autoref{fig:comparison_phot_astrom_par_umag}. BAGLE and pyLIMA agree well, while VBMicrolensing and BAGLE disagree significantly by more than 1$\times\thetaE$. 

\begin{figure*}
   \centering
    \includegraphics[width=0.85\textwidth]{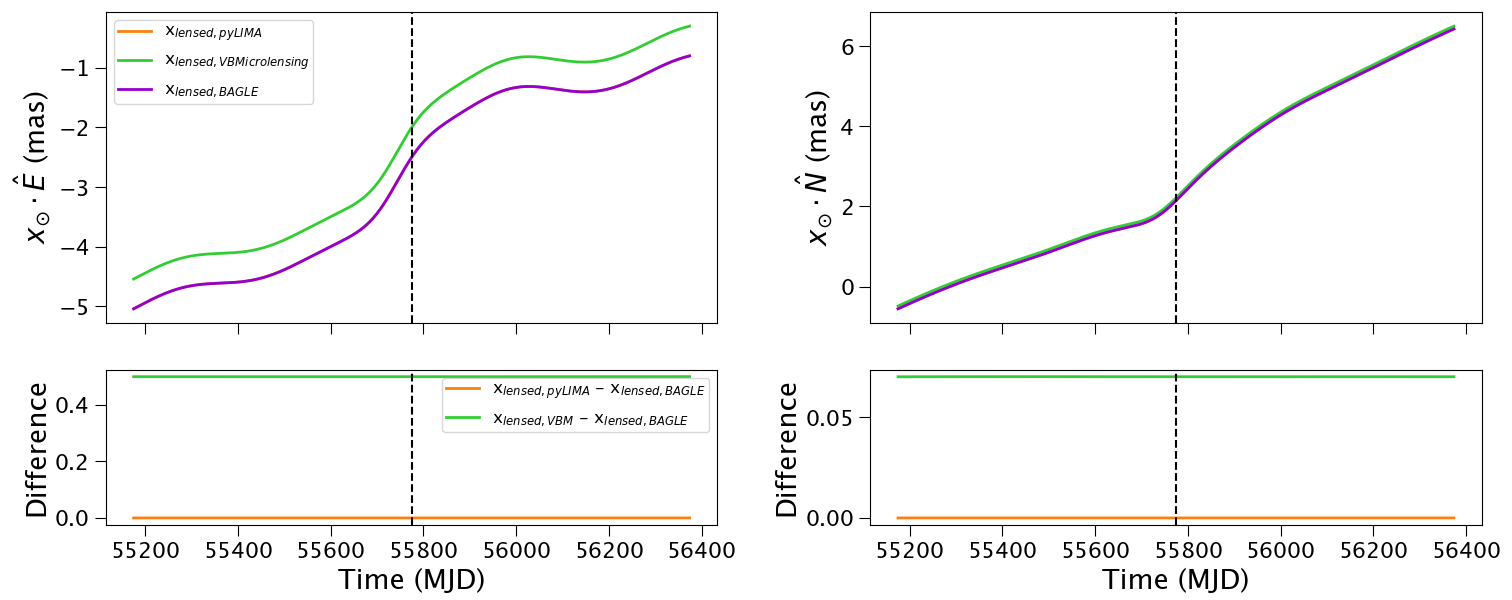}
    \caption{Comparison between BAGLE, pyLIMA, and VBMicrolensing's lensed image centroids for a simulated PSPL event with parallax. The heliocentric model parameters for this event are
    $t_{E,\odot}=100$ days, $u_{0,\odot}=1.0$, $t_{0,\odot}=55775$ MJD, $\pi_{E,E,\odot}=0.2$, $\pi_{E,N,\odot}=0.1$, $b_{sff}=1$, $mag_S=19$ mag, $\alpha_L=269.3$ deg, $\delta_L=-30.4$ deg
    $\pi_S=0.1$ mas, $\thetaE=3.0$ mas, $x_{S,0,E,\odot}=-3.0$ mas, $x_{S,0,N,\odot}=3.0$ mas, $\mu_{S,E,\odot}=1.0$ mas yr$^{-1}$, and $\mu_{S,N,\odot}=2.0$ mas yr$^{-1}$.
    ({\em Top}) Unresolved lensed image centroid in the East ({\em left}) and North ({\emph right}) directions. 
    ({\em Bottom}) Centroid difference between pyLIMA or VBMicrolensing and BAGLE models.
    }
    \label{fig:comparison_phot_astrom_par_lensed}
\end{figure*}

We also compare the actual lensed image centroid between BAGLE, VBMicrolensing and pyLIMA in Figure \ref{fig:comparison_phot_astrom_par_lensed}. 
We find that VBMicrolensing and BAGLE disagree by a constant offset. We attribute this offset to the fact that VBMicrolensing does not input the initial position of the source. While it does assume that the lens is at the origin at $\tnotgeotr$, this choice seems un-physical and requires a transformation to pyLIMA and BAGLE's initial position quantified by the offset in Figure \ref{fig:comparison_phot_astrom_par_lensed}.

\begin{figure}[!h]
   \centering
    \includegraphics[width=0.48\textwidth]{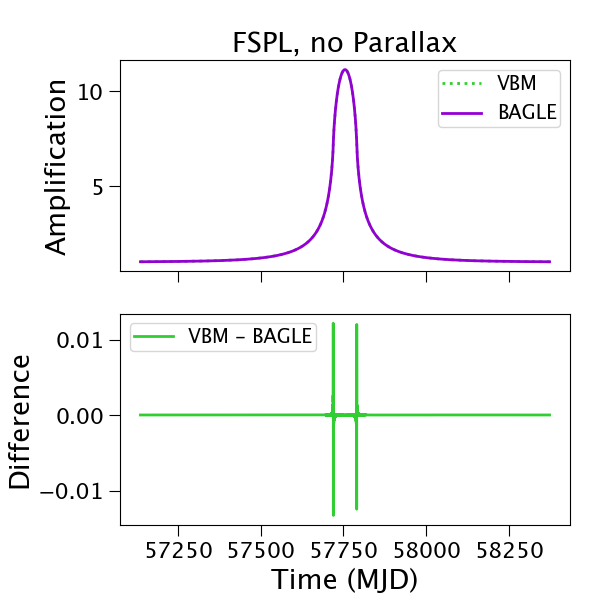}
    \caption{Comparison between BAGLE and VBMicrolensing's amplifications for a simulated FSPL event without parallax. The amplification \emph{(top)} and residuals with respect to the BAGLE model \emph{(bottom)} are shown over time. The heliocentric model parameters for this event are
    $\tnot = 55775$ MJD, $\uo = 0.044$, $\tE = 206$ days, $\bsff = 1$, and $\rho_* = 0.177 \, \theta_E$
    }
    \label{fig:phot_compar_fspl}
\end{figure}

Lastly, we compare amplifications for FSPL models between VBMicrolensing and BAGLE.  To calculate the FSPL lightcurve amplification, VBMicrolensing applies a quadrupole correction to the point-source, point-lens amplification. Furthermore, the elliptical integrals dependent on the source radius $\rho_*$ that are used to calculate magnification are pre-calculated in a file labelled ``ESPL.tbl''' and loaded during the call to VBMicrolensing. This approach works optimally for cases where the source is far away from the lens caustic. Unlike VBMicrolensing, BAGLE performs contour integration on the image area. We compare the BAGLE and VBMicrolensing lightcurves and present our results in  \autoref{fig:phot_compar_fspl}. We find that the two packages yield similar light curves with residuals at an order of magnitude of $10^{-2}$ for high magnification events.

\section{Validation of Model Fitter}
\label{sec:validate_fitter}

\begin{figure*}[htpb]
    \centering
    \includegraphics[width=0.9\textwidth]{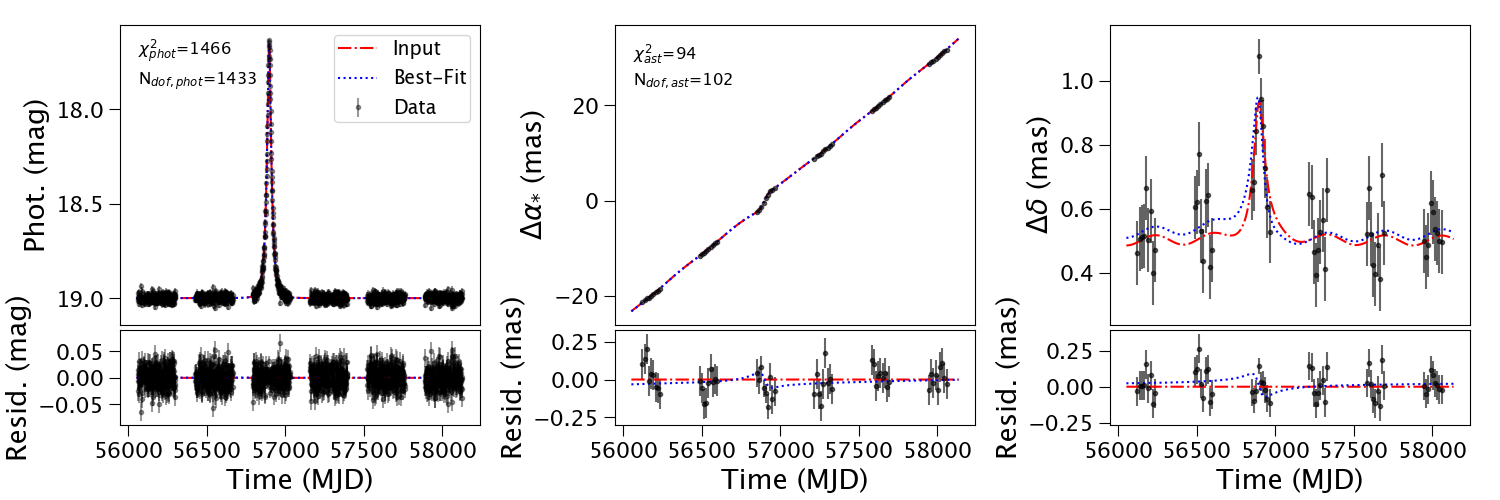}
    \caption{
    Mock photometric ({\em left}) and astrometric ({\em middle, right}) data are shown in {\em black} for a bulge microlensing event with input parameters from Table \ref{tab:pspl_fit_params}. The input model is shown in {\em red}. The best-fit model 
    is shown in {\em blue}. 
    Residuals with respect to the input model are shown at the {\em bottom}.
    \label{fig:pspl_fit}
    }
\end{figure*}

To test the efficacy of \bagle's model fitting procedures, we simulate mock lightcurves with photometric and/or astrometric noise, fit them with \bagle, and compare the input and output parameters.
The mock data was generated using the \bagle~ \texttt{PSPL\_PhotAstrom\_Par\_Param2} model class with input parameters shown in Table \ref{tab:pspl_fit_params}. 
Data points were sampled from the model over a 2000 day window with a cadence of 1 day and 14 days for photometry and astrometry, respectively.
Seasonal gaps were included as appropriate for bulge visibility from OGLE (photometry) and Keck (astrometry). 
Random noise was added assuming a photometric SNR=63 ($\sigma_{mag}$=0.016) and an astrometric error of 0.15 mas at mag=19.
The observed lightcurve and trajectory are plotted with the input model and the best-fit model, which are in good agreement (Figure \ref{fig:pspl_fit}). 
The best-fit parameters are shown in Table \ref{tab:pspl_fit_params} and also match the input values, within uncertainties. 

% Refresh model and fit:
% lens_math.run_model_fitter(run_fit=True)
% Re-plot without refitting
% lens_math.run_model_fitter(run_fit=False)

% Code to make table data:
% lens_math.make_model_fitter_table()
\begin{deluxetable}{lrr}
\tabletypesize{\footnotesize}
%\tablewidth{0pt}
\tablecaption{BAGLE Model Input and Best-Fit Parameters \label{tab:pspl_fit_params}}
\tablehead{
\colhead{Parameter} & 
\colhead{Input} & 
\colhead{Output}
}
\startdata
\multicolumn{3}{l}{\textbf{Fit Parameters}} \\ 
$t_0$ (MJD) & 56900.00 & 56900.60 $\pm$ 1.07 \\ 
$u_0$ & 0.30 & 0.30 $\pm$ 0.01 \\ 
$t_E$ (days) & 30.0 & 30.1 $\pm$ 0.3 \\ 
$\theta_E$ (mas) & 3.00 & 2.87 $\pm$ 0.19 \\ 
$\pi_{E,E}$ & 0.050 & 0.068 $\pm$ 0.037 \\ 
$\pi_{E,N}$ & 0.000 & -0.006 $\pm$ 0.008 \\ 
$b_{SFF,I}$ & 1.000 & 1.000 $\pm$ 0.005 \\ 
$I_{src}$ (mag) & 19.000 & 19.000 $\pm$ 0.006 \\ 
$\pi_S$ (mas) & 0.125 & 0.125 $\pm$ 0.052 \\ 
$x_{S0,\alpha*}$ (mas) & 0.00 & -0.00 $\pm$ 0.00 \\ 
$x_{S0,\delta}$ (mas) & 0.00 & 0.00 $\pm$ 0.00 \\ 
$\mu_{S,\alpha*}$ (mas/yr) & 10.00 & 10.01 $\pm$ 0.18 \\ 
$\mu_{S,\delta}$ (mas/yr) & 0.00 & 0.00 $\pm$ 0.03 \\ 
\tableline
\multicolumn{3}{l}{\textbf{Fixed Parameters}} \\ 
$\alpha$ (hr)& 259.5 & \\ 
$\delta$ (deg)& -29.0 & \\ 
Location& earth & \\ 
\tableline
\multicolumn{3}{l}{\textbf{Derived Parameters}} \\ 
$M_L$ ($\msun$) & 7.4 & 8.3 $\pm$ 10.6 \\ 
$\pi_L$ (mas) & 0.275 & 0.321 $\pm$ 0.119 \\ 
$\mu_{L,\alpha*}$ (mas/yr) & -26.52 & -24.57 $\pm$ 2.29 \\ 
$\mu_{L,\delta}$ (mas/yr) & 0.00 & 2.33 $\pm$ 2.63 \\ 
$\mu_{rel,\alpha*}$ (mas/yr) & 36.52 & 34.58 $\pm$ 2.28 \\ 
$\mu_{rel,\delta}$ (mas/yr) & 0.00 & -2.33 $\pm$ 2.62 
\enddata
\end{deluxetable}

\section{Results}
\label{sec:results}

The \bagle~ package allows us to explore many aspects of microlensing events. 
In this section, we investigate
\begin{itemize}
    \setlength\itemsep{0em}
    \item \S\ref{sec:results_mass} - Dependency on lens mass.
    \item \S\ref{sec:results_parallax} - Dependency on parallax from Earth or a satellite.
    %\item \S\ref{sec:results_degeneracies} - Degeneracy between $t_E$, $u_0$, and $b_{sff}$.
    \item \S\ref{sec:results_lumlens} - Impact of luminous Lens.
    \item \S\ref{sec:results_model_runtimes} - Model-generation run times.
    \item \S\ref{sec:results_fit_runtimes} - Model-fit run times.
\end{itemize}

\subsection{Results: Lens Mass}
\label{sec:results_mass}

\begin{figure}
    \centering
    \includegraphics[width=0.5\textwidth]{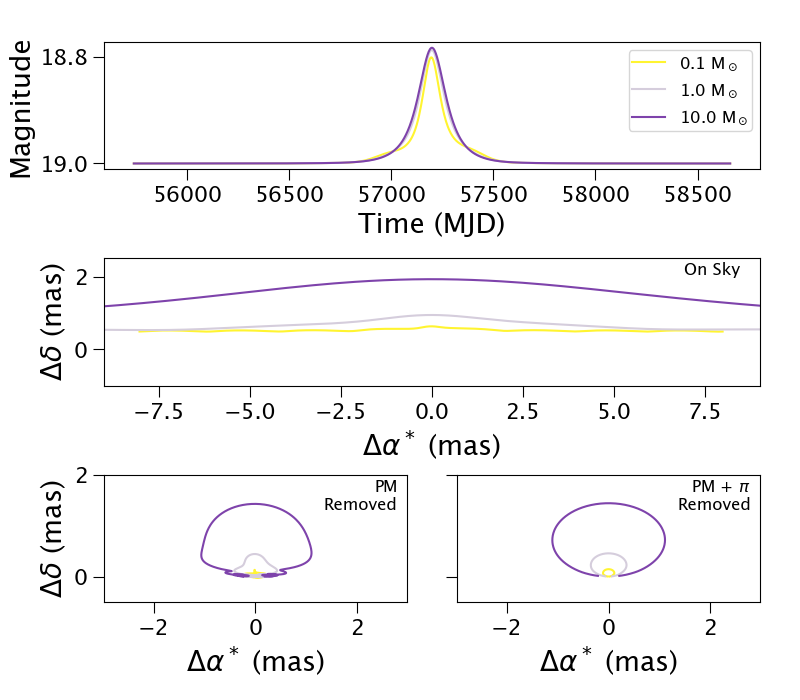}
    \caption{The impact of changing lens mass in \bagle~ models for PSPL photometry ({\em top}) and astrometry ({\em middle}). Three models are shown ({\em color}) with similar lightcurves; but with different lens masses and $\mu_S \propto 2.0 (M_L / 0.1 M_\odot)^{0.5}$ mas yr$^{-1}$ and $\beta \propto 0.5 (M_L / 0.1 M_\odot)^{0.5}$.
    Remaining model parameters include $\Xsovec$ = [0.0, 0.5] mas, $\mulvec$ = [0, 0] mas/yr, $d_L$ = 3 kpc, $d_S$ = 8 kpc, $b_{SFF}$ = 1, $mag_S$ = 19 mag, $t_0$ = 57200 MJD, $\alpha$ = 262.5$\degr$, $\delta$ = -30$\degr$.    
    }
    \label{fig:vs_mass}
\end{figure}

One of the most powerful uses of gravitational lensing is to determine the lens mass, even if the lens is dark or too faint to detect. 
The impact of the lens mass is most evident in the astrometric trajectory on sky. 
Figure \ref{fig:vs_mass} shows both the astrometry and photometry for different lens masses in the case where the light curves are almost identical. 
Since the lightcurve shape duration is set primarily by $\tE = \thetaE / \murel$, we scale $\murel = \mu_{S}$ with lens mass as $\mu_S \propto (M_L/0.1 \msun)^{0.5}$ mas yr$^{-1}$. 
Figure \ref{fig:vs_mass} shows that very similar lightcurves produce very different astrometric trajectories on the sky and that more massive lenses have stronger astrometric microlensing signals. 

\subsection{Results: Parallax}
\label{sec:results_parallax}

\begin{figure}
    \centering
    \includegraphics[width=0.5\textwidth]{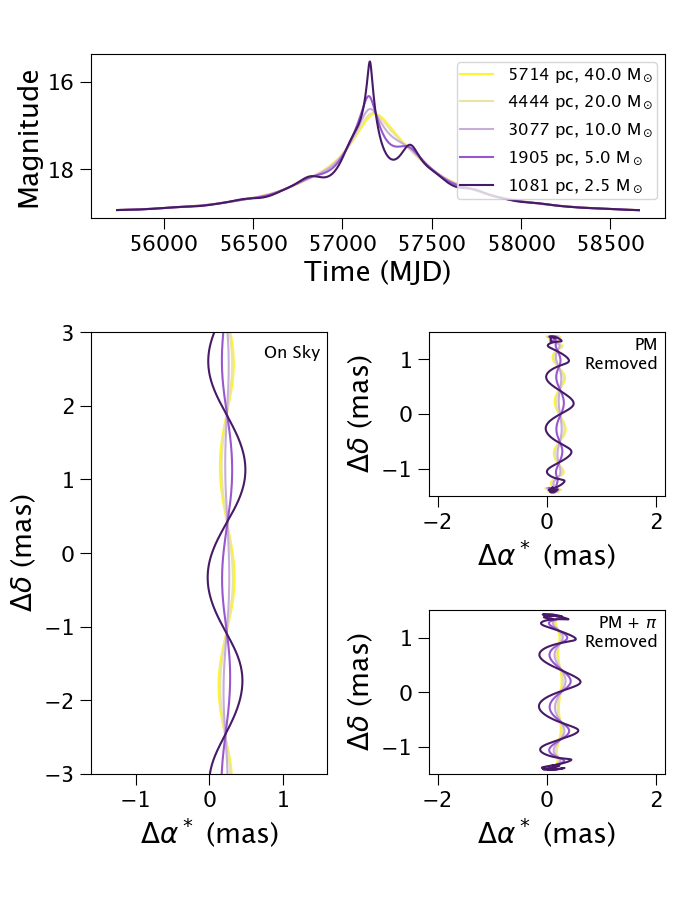}
    \caption{
    The impact of changing the lens parallax in \bagle~ models for PSPL photometry ({\em top}) and astrometry ({\em bottom}). Five models are shown ({\em color}) with similar lightcurves; but with different lens distances and lens masses $M_L \propto 10 M_\odot (0.2 / \pi_{rel})$.
    Remaining model parameters include $\beta = 0.5$, $\Xsovec$ = [0.00, 0.50] mas, $\mulvec$ = [0, 0] mas/yr, $\musvec$ = [0, 2] mas/yr, $d_S$ = 8 kpc, $b_{SFF}$ = 1, $mag_S$ = 19 mag, $t_0$ = 57200 MJD, $\alpha$ = 262.5$\degr$, $\delta$ = -30$\degr$.    
    }
    \label{fig:vs_parallax}
\end{figure}

Microlensing lightcurves are remarkably simple in the absence of parallax: single-peaked and symmetric. 
However, realistic lightcurves are multi-peaked as shown in Figure \ref{fig:vs_parallax} where the parallax is increased with all other parameters, including $\tE$ being equal.
Note that to maintain a constant $\tE$ and overall lightcurve shape, the lens mass must scale as $M_L \propto (1 / \pi_{rel})$ (top panels). 
As a result, lightcurves with similar $\tE$ and different $\piE$ lead to very different lensed astrometry on the sky (bottom panels).

Parallax effects are most evident in long-duration events ($\tE>75$ days for $\piE>0.15$) for typical ground-based photometric precisions of $\sim 5\%$;
however, space-based photometric precisions of $<$1\% reveal detectable parallax signals in events with $\tE>50$ days for $\piE>0.05$. 
As a demonstration, mock lightcurves are generated using a \texttt{PSPL\_Phot\_Par\_Param1} \bagle~ models under two different observing scenarios:
\begin{itemize}
\item {\em Mock OGLE:} Data points are generated assuming SNR=63 at I=19 mag over 5 years with daily cadence and bulge visibility over the year typical for the OGLE telescope.
\item {\em Mock Roman:} Data points are generated assuming SNR=1 at F146W=28 mag over 5 years with $\sim$15 min cadence during 6 fast visibility-seasons and 10 day cadence during 4 slow visibility-seasons, as described in \citet{ROTAC:2025}.
\end{itemize}
Lightcurves are generated for events over a grid of $\tE$ (range=1-200 days, step=2 days) and $\piE$ (range=0.005-0.30, step=0.005) and with $\tnot=61693$ MJD, $\uo=0.3$, $\phi_{\mu_{rel}}=45\degr$, $b_{sff,OGLE}=1$, $b_{sff,Roman}=1$, $\alpha_L=259.5\degr$, $\delta_L=-29\degr$. 
The unlensed source brightness in Vega-magnitudes was taken as $mag_{S,OGLE}=20$ mag, $mag_{S,Roman}=18.5$ mag as would be typical for reddened bulge stars. 
The lensed photometry was perturbed by the expected photometric uncertainties for each time step. 
To estimate how sensitive each survey is to $\piE$, the mock lightcurves were used to calculate the Fisher information matrix, resulting covariance matrix and the predicted uncertainty on $\piE$.

Figure \ref{fig:piE_sensitivity} shows the resulting signal-to-noise, SNR = $\piE / \sigma_{\pi_E}$ over the grid of simulated lightcurves. 
The Roman lightcurves will result in significantly better measurements of the microlensing parallax.
In particular, Roman can detect microlensing parallax signals from photometry alone for events that are twice as short in duration ($\tE$) as from ground-based photometric surveys, such as OGLE. 
Roman is also far more sensitive to small $\piE$ events, which is particularly important for discovery of isolated black hole lenses \citep{Lam:2020}. 

\begin{figure*}
    \centering
    \includegraphics[width=\textwidth]{figures_plot_snr_piE_ogle_roman.png}
    \caption{
    Predicted signal-to-noise ratio (SNR) on the measurement of $\piE$ based on {\em Mock OGLE} (left) and {\em Mock Roman} (right) lightcurves. 
    Detection thresholds of $3\sigma$, $6\sigma$, and $10\sigma$ are over-plotted in white. 
    Roman will detect $\piE$ from lightcurves with 2$\times$ shorter duration than OGLE and other ground-based photometric surveys.
    \label{fig:piE_sensitivity}
    }
\end{figure*}

\subsection{Results: Luminous Lenses}
\label{sec:results_lumlens}

\begin{figure*}
    \centering
    \includegraphics[width=0.75\textwidth]{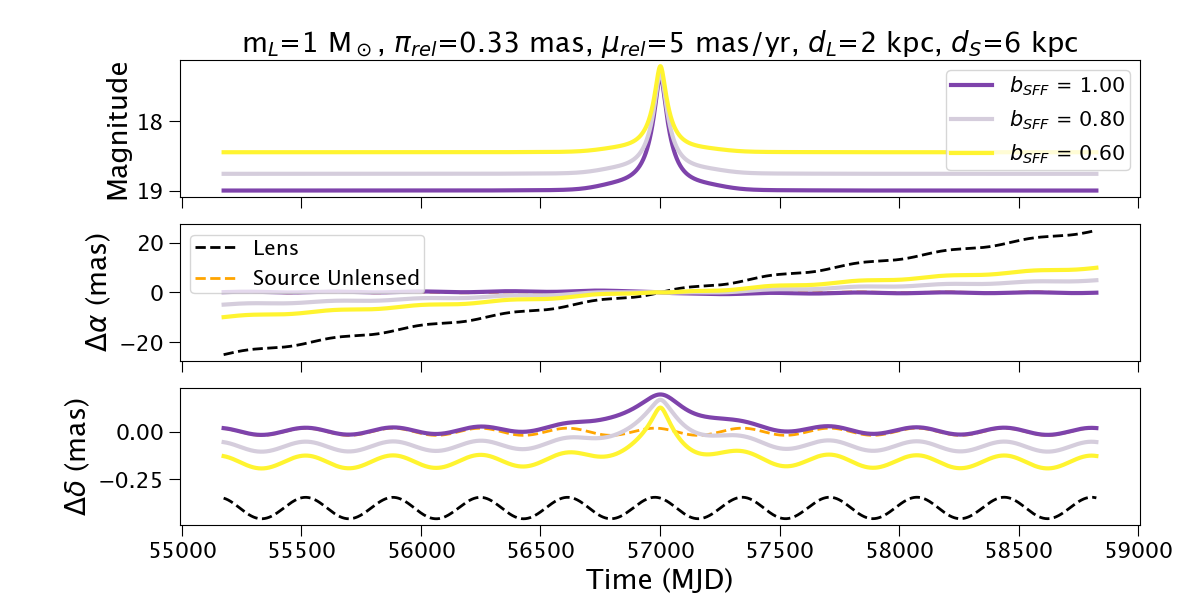}
    \caption{The impact of blending on the photometry (top panel) and astrometry (middle, bottom panels) for a microlensing event. 
    In this case, all blending is assumed to come from a luminous lens. 
    The complete set of model parameters for the \texttt{PSPL\_PhotAstrom\_Par\_Param1} model shown is 
    $M_L=1 \msun$, $\tnot=57000$ MJD, $\Xsovec=[0'',0'']$,
    $\beta=0.4$ mas, $\vec{\mu_S}=[0,0]$ mas yr$^{-1}$,
    $\vec{\mu}_L=[5,0]$ mas yr$^{-1}$, 
    $d_L=2$ kpc, $d_S=6$ kpc, $\alpha_L=262.5\degr$, $\delta_L=-29\degr$.
    }
    \label{fig:vs_bsff}
\end{figure*}

Luminous lenses can significantly complicate both the photometric and astrometric signal. 
For photometry, blended light raises the background base flux level and can even prevent detection of the microlensing event except at the very peak. 
This is especially relevant in ground-based, seeing-limited photometric surveys with spatial resolutions of $\sim$1" where there are typically multiple stars within a 1" patch at visible wavelengths towards the Galactic bulge. 
Figure \ref{fig:vs_bsff} (top panel) shows the impact of different amounts of blending, $\bsff$, on photometric lightcurves. 

Astrometric trajectories are impacted by blending, even in the absence of gravitational lensing. 
Figure \ref{fig:vs_bsff} (middle, bottom panels) shows the impact of blending on the observed trajectory.
Even well before the microlensing event occurs, the observed position is the flux-weighted centroid of the light from both the source and lens (and possible luminous companions). 
If such a trajectory were fit with a typical motion model with proper motion and parallax, the resulting proper motion and parallax would not reflect that of either the source or lens. 
Thus an observer must take care to account for blending in deriving baseline quantities well before and after the lensing event.
As long as the source and lens remain unresolved, both must be modeled in baseline using the $\bsff$ derived during the microlensing event.

\begin{figure}
    \centering
    \includegraphics[width=0.48\textwidth]{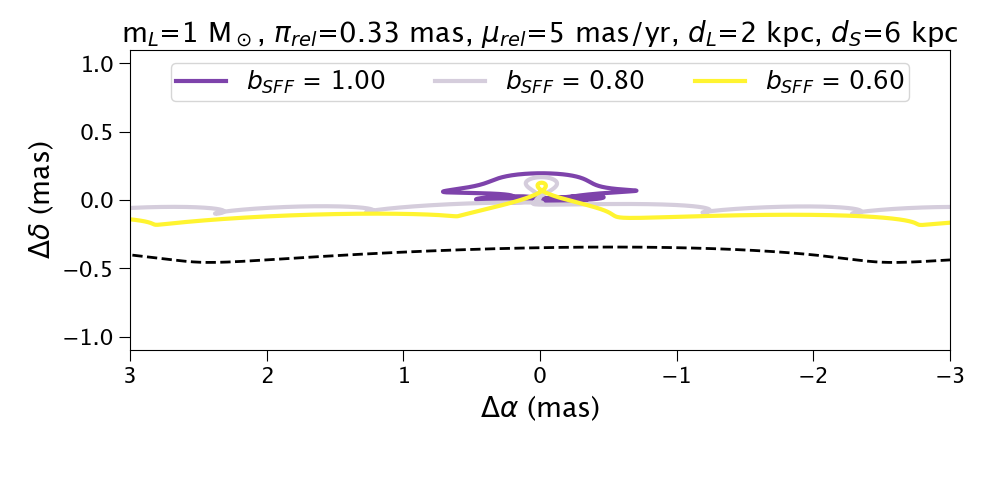}
    \caption{
    The observed trajectory on the sky is significantly impacted by blending as shown in the trajectories shown in different colors for different $\bsff$. 
    The lens trajectory is shown in the black dashed line. 
    In this event, the source has no proper motion; but does have parallax. 
    \label{fig:on_sky_vs_bsff}
    }
\end{figure}

Figure \ref{fig:on_sky_vs_bsff} shows how the observed trajectory on the sky becomes even more complex with the addition of lensing. 
For especially long-duration events, where the annual parallax cycle is repeated multiple times during the event (i.e. $t_E > 2$ years), the parallax signal seen during the event is dramatically impacted and is not representative of either the lens or source parallax.

\subsection{Results: Model Run Times}
\label{sec:results_model_runtimes}

We compare the computational run time of \bagle~ with those of other microlensing packages. 
Runtimes are calculated for models by generating mock data for an event with $t_E=100$ days sampling 2000 time steps over 5.5 years. 
The time tests are repeated 100 times and the mean and standard deviation are recorded for test.
Tests were performed with \bagle, VBMicrolensing v5.3.3, pyLIMA v1.9.8, and MulensModel v3.3.1. 
All tests were conducted on a 2021 iMac equipped with a M1 processor using a python 3.11 environment and the ipython kernel.
Note that tests run in an ipython kernel were significantly faster than those run in a Jupyter notebook.

Tests were first performed using a PSPL photometry-only model with no parallax. 
Run times differ when including the instantiation of the model object or only computing amplifications on an existing instance. 
Both results are presented in Table \ref{tab:runtimes_nopar}.
The full runtime is applicable to large-scale simulations of multiple events located at different sky coordinates.
In this case, BAGLE and VBMicrolensing generate mock lightcurves the fastest, while pyLIMA has a large initial overhead for each mock event.
The pre-instantiated runtime is applicable to running many different sets of model parameters for an event at the same sky location (i.e.~model fitting). 
In this case, all four packages are quick for no-parallax model generation. 

\begin{deluxetable}{lcc}
\tablecaption{PSPL Model-Generation Runtimes for Photometry-Only Models Without Parallax \label{tab:runtimes_nopar}}
\tablehead{
  \colhead{} &
  \multicolumn{2}{c}{\textbf{Runtime (ms)}} \\
  \colhead{\textbf{Model}} &
  \colhead{\textbf{Full}} &
  \colhead{\textbf{Pre-Instantiated}} \\
}
\startdata
BAGLE           &    0.066 $\pm$    0.008  &    0.049 $\pm$    0.005 \\
pyLIMA          &   12.76 $\pm$    1.02  &    0.20 $\pm$    0.02 \\
MulensModel     &    1.03 $\pm$    0.71  &    0.24 $\pm$    0.01 \\
VBMicrolensing  &    0.109 $\pm$    0.003  &    0.113 $\pm$    0.002
\enddata
\end{deluxetable}

Runtimes using PSPL photometry-only models with parallax are significantly longer due to the need to calculate Earth's position relative to the Solar System Barycenter over time. 
Table \ref{tab:runtimes_par} shows the runtimes for the parallax case is 6 times longer for \bagle~ in the pre-instantiated case, which is due to our use of the full JPL ephemeris to set Earth's location in order to accurately capture the influence of the Moon and other Solar System bodies. 

In contrast, VBMicrolensing uses a lookup table derived from approximate orbital elements for Earth's orbit, which is less accurate, but much quicker to evaluate once loaded up. 
This lookup table approach explains why the instantiation time for VBMicrolensing is significantly longer. 
The origin of the long instantiation time for pyLIMA is unclear as, like \bagle, pyLIMA uses the \texttt{astropy.coordinates} package to compute Earth's location, although pyLIMA uses the {\em builtin} analytic ephemeris rather than the JPL ephemeris by default. 
For simulations of many events, the fast \bagle~runtime is a significant advantage. For model-fitting, VBMicrolensing will be the fastest to run, although the parallax model will be less accurate compared to other packages.

\begin{deluxetable*}{lcccc}
\tablecaption{PSPL Model-Generation Runtimes for Models With Parallax \label{tab:runtimes_par}}
\tablehead{
  \colhead{} & 
  \multicolumn{4}{c}{{\textbf{Runtime (ms)}}} \\
  \colhead{} & 
  \multicolumn{2}{c}{\textbf{Photometry-Only}\hspace{0.5in}} &
  \multicolumn{2}{c}{\textbf{Photometry+Astrometry}\hspace{0.2in}} \\
  \colhead{\textbf{Model}} &
  \colhead{\textbf{Full}} & 
  \colhead{\textbf{Pre-Instantiated}} & 
  \colhead{\textbf{Full}} & 
  \colhead{\textbf{Pre-Instantiated}}
}
\startdata
BAGLE           &    0.32 $\pm$    0.03  &    0.29 $\pm$    0.18  &    0.87 $\pm$    0.17  &    0.76 $\pm$    0.36 \\
pyLIMA          &  57.07 $\pm$   33.32  &    0.20 $\pm$    0.05  &  106.91 $\pm$   31.91  &    0.50 $\pm$    0.12 \\
MulensModel     &    2.11 $\pm$    0.34  &    0.31 $\pm$    0.16 & - & - \\
VBMicrolensing  &   8.57 $\pm$   7.63  &    0.11 $\pm$    0.01  &  129.12 $\pm$  8.43  &    0.57 $\pm$    1.54 \\
\enddata
\end{deluxetable*}

We calculate runtimes for models with parallax and generating both photometry and astrometry. 
In this test, MulensModel is not included as it does not currently generate astrometry. 
Table \ref{tab:runtimes_par} shows the results. \bagle, pyLIMA, and VBMicrolensing have statistically consistent run times for pre-instantiated model generation. 

Finally, runtimes were also compared for FSPL models in \bagle~ and VBMicrolensing. 
For this test, models were evaluated at 210 timesteps of 2.83 years. 
Time tests were run over 100 iterations to estimate uncertainties on the runtimes. 
The FSPL runtime results are presented in \autoref{tab:runtimes_nopar_fspl}. 
We find that the numerical approach in \bagle~is much slower than the calculations performed in VBMicrolensing; 
however, the \bagle~approach is likely more accurate as described in \S\ref{sec:fspl}. 

\begin{deluxetable}{lcc}
\tablecaption{FSPL Model-Generation Runtimes for Photometry-Only Models Without Parallax
\label{tab:runtimes_nopar_fspl}}
\tablehead{
  \colhead{}
   & 
  \multicolumn{2}{c}{{\textbf{Runtime (ms)}}} \\
  \colhead{\textbf{Model}} &
  \colhead{\textbf{Full}} &
  \colhead{\textbf{Pre-Instantiated}}
}
\startdata
BAGLE           &    60.498 $\pm$ 0.293  &  61.085 $\pm$  0.513 \\
VBMicrolensing  &    0.052  $\pm$ 0.027  &   0.020 $\pm$  0.012 \\
\enddata
\end{deluxetable}

\subsection{Results: Fit Run Times}
\label{sec:results_fit_runtimes}

We compare the time required to run a fit with BAGLE, VBMicrolensing+RTModel, pyLIMA, and MulensModel on a photometry-only mock light curve, generated using parameters presented in Table \ref{tab:pspl_fit_params}.
The mock photometric data that were fit are shown in Figure \ref{fig:pspl_fit} (left panel) with input parameters from Table \ref{tab:pspl_fit_params}.

In the comparison, we used the MultiNest nested sampling fitting routine with BAGLE and MuLens, the Markov Chain Monte Carlo (MCMC) fitting routine with pyLIMA, and the Levenberg-Marquardt (LM) algorithm with RTModel. 
For BAGLE fits, the MultiNest parameter for the number of live points was set to 350. 
RTModel is the wrap-around package that fits VBMicrolensing models to data and is described in \citet{BozzaRT:2024}.
Note that RTModel does not have a point-source, point-lens model available for fitting. We fit it with a finite-source, point-lens model that places no constraints on the source radius $\rho_*$.

\begin{deluxetable}{cccc}
\tabletypesize{\footnotesize}
%\tablewidth{0pt}
\tablecaption{Model-Fit Runtimes for Photometry-Only with Parallax  \label{tab:runtimes_fit_par} }
\tablehead{
\colhead{\textbf{Model}} & 
\colhead{Runtime (s)} & 
\colhead{$\bar{\chi}^2$} & 
\colhead{Notes}
}
\startdata
BAGLE & 6.97 & 0.95 & MultiNest Nested Sampling\\ 
RTModel & 9.48 & 0.96 & Levenberg–Marquardt Fitting\\ 
pyLIMA & 18.48 & 0.96 & MCMC Fitting\\ 
MuLens & 8.72 & 0.96 & MultiNest Nested Sampling \\ 
\enddata
\tablenote{A comparison of fitting runtimes for different photometry-only models with parallax to fit the photometry of a mock event.
}
\end{deluxetable}

The tests were conducted on a 2021 iMac equipped with 8 M1 processors. By default, RTModel uses the maximum number of available processors to run fits in parallel.  Both RTModel and BAGLE completed the fits with 8 processors. On the other hand, while pyLIMA applies the multiprocessing library to parallelize aspects of its model fitting processes, only a single processor was used in the call to pyLIMA's MCMC fitting routine. It was unclear if MuLensModel parallelizes by default. A single processor was used to run the fit using MuLensModel. 

We present the fitting runtimes in \autoref{tab:runtimes_fit_par}. All packages in our comparison give similar reduced chi-squared $\bar{\chi}^2$ value.
We find that BAGLE produces comparable runtimes to RTModel and MuLensModel for this photometry-only model fit. 
\bagle's use of nested sampling (along with MuLensModel) produces a full posterior probability distribution for the parameters and allows for easy hypothesis testing when comparing different model classes. 
Further comparisons of the accuracy of the model fits and derived parameters will be conducted in the future. 

\section{Conclusion \& Future Directions}
\label{sec:conclusion}

We have presented the BAGLE microlensing event modeling and fitting package, which jointly models and fits photometry and astrometry. This paper presents the point-source, point-lens and uniform-source, point-lens model implementations.
The BAGLE implementations of Gaussian processes and binary source and lens models are beyond the scope of this paper and will be presented in
\citet{Chen:2025} and \citet{Bhadra:2025}, respectively.

We presented the model parameters, equations, and fitting procedures and their relationship to alternative parameterizations in the literature.
Notably, \bagle~ works in a coordinate system anchored to the Solar System Barycenter, which allows for robust estimation of astrometric trajectories on sky in addition to the photometric lightcurves that most other packages produce. 
The lightcurves and trajectories output from \bagle~ are in broad agreement with other packages, although \bagle's parallax model is more accurate thanks to reliance on JPL ephemerids for Earth's position. 
In parallax cases, the differences between lightcurves across the packages are large enough, in some cases, to be detectable in Roman GBTDS observations.

\bagle's model-fitting runtimes are comparable to VBMicrolensing/RTModel and MulensModel on a single event and faster than pyLIMA.
Simulation of large numbers of events at different sky coordinates is $5-400\times$ faster with \bagle for all point-source, point-lens models.
In the future, we plan to implement automatic differentiation to increase \bagle's fitting efficiency and allow for the use of GPU-accelerated inference packages such as PyTorch or TensorFlow. 

Thus far, the \bagle~ joint photometric-astrometric fitting capabilities have been published for a relatively small set of ground-based photometric light curves with slow-cadence (a couple of points per year), ground-based adaptive optics or space-based astrometric follow-up. 
The package is now used routinely to simulate microlensing lightcurves for Rubin and Roman \citep{Abrams:2025rubin,GBTDS_DefCom:2025}. 
With the Roman Space Telescope's launch approaching, \bagle's robust support for photometry and astrometry will be critical to leverage the influx of simultaneous, high-cadence photometric and astrometric microlensing data and study the properties and prevalence of isolated stellar mass black holes.

\begin{acknowledgments}
    We thank David Bennett, Valerio Bozza, Etienne Bachelet, and Radek Poleski for comments on package comparisons. We thank the Roman Galactic Exoplanet Project Infrastructure Team (RGES-PIT) working groups on microlens modeling and astrometry for useful discussions.
    The authors acknowledge support from the National Science Foundation under grant No.~2108185, the Heising-Simons Foundation under grant No.~2022-3542, and the Association of Universities for Research in Astronomy Space Telescope Science Institute under grant No.~HST-GO-17081.004-A and HST-GO-16658.001-A.
\end{acknowledgments}

\software{Numpy \citep{oliphant2006guide}, Matplotlib \citep{Hunter:2007}, Astropy \citep{astropy:2013,astropy:2018,astropy:2022}, SciPy \citep{2020SciPy-NMeth}}

\appendix
\section{BAGLE Development}\label{sec:bagle_dev}

\subsection{Making a New Model}
Each model is, as described in Section \ref{sec:bagle_implemenation}, constructed by combining parent classes that contain the desired features for the model. Each
model must have one class from each required class family. In addition to this, there
are several rules that must be followed when creating a new class.

\begin{enumerate}
    \item The data class must match the parameterization class. For example,
        if the chosen data class is \texttt{PSPL\_Phot}, then the parameter class
        must be \texttt{PSPL\_PhotParam$<$N$>$} for some valid parameterization number \texttt{N}. %If the data class is `PSPL\_PhotAstrom`, then the parameter
        %class must be one of the classes with a PhotAstromParam.

    \item Models are built using python's multiple inheritance feature. Therefore
        the order in which the parent classes are listed in the model class'
        definition matters. Parent classes to models should always be listed
        in the order:
        \begin{enumerate}
            \item ModelClassABC
            \item Data Class
            \item Parallax Class
            \item Parameterization Class
        \end{enumerate}
        If using the optional GP class, then the order is
        \begin{enumerate}
            \item ModelClassABC
            \item GP Class
            \item Data Class
            \item Parallax Class
            \item Parameterization Class
        \end{enumerate}
    \item  Each class must be given the \texttt{@inheritdocstring} decorator, and include
        the following three commands in the model's \texttt{\_\_init\_\_}:
        \begin{itemize}
            \item \texttt{super().\_\_init\_\_(*args, **kwargs)}: Inherit the \texttt{\_\_init\_\_} from the Parameterization Class.
            \item \texttt{startbases(self)}: Run a \texttt{start} command on each parent class, giving each parent class a chance to run a set of functions upon instantiation.
            \item \texttt{checkconflicts(self)}: Check to confirm that the combination of parent classes in the model are valid.
        \end{itemize}
    \item  Models should be named to reflect the parents classes used to construct
        it, as outlined in the above sections.

\end{enumerate}

Note:
All times must be reported in MJD.

\subsection{Other Fitters}
PyMultiNest \citep{pymultinest} is currently the only fitter native to BAGLE. However, other fitters may be run on the existing BAGLE models. 
Fitters that support user-supplied likelihood functions are the most straight-forward to use as the likelihood is implemented as a function on the {\tt bagle.model\_fitter.MicrolensSolver} class.
Limited tests have been run using Dynesty [CITE] and PyMC \citep{PyMC} and further work is in progress.

\section{Geocentric($t_r$) Coordinate Frames}
\label{sec:helio_geo_convert}

\subsection{Heliocentric  $\rightarrow$ Geocentric($t_r$)}

The photometric parameters that are effected by transforming the reference frame from heliocentric to geocentric($t_r$) include $\tnot$ and $\tE$.
These are converted into
$\tnotgeotr$ and $\tEgeotr$ in the following manner. 
In order to convert the Einstein crossing time, we use
Eq.~\ref{eq:murel_helio_geotr_conv} and 
\begin{eqnarray}
    \thetaE &= \tE \murel = \tEgeotr \murelgeotr
\end{eqnarray}
which gives
\begin{eqnarray}
    \tEgeotr &= \tE \frac{\murel}{|\murelvec - \pirel \vect{\dot{P}}(t_r)|}.
\end{eqnarray}

In photometry-only fits, we don't have access to $\murel$ and $\pirel$ information; but instead have $\murelhat = \piEhat$ and $\piE$. 
Since 
\begin{align}
    \frac{\murelvec}{\pirel} &= \frac{\murelhat}{\piE \tE} \\
    \frac{\murelvecgeotr}{\pirel} &= \frac{\murelhatgeotr}{\piE \tEgeotr} 
\end{align}
the Einstein crossing time can be expressed as
\begin{align}
\tEgeotr &= \tE \frac{ \frac{1}{\piE \tE} }{\left| \frac{\murelhat}{\piE \tE} - \vect{\dot{P}}(t_r) \right|} \\
\Aboxed{ \tEgeotr &= \tE \frac{ 1 }{\left| \murelhat - \piE \tE \vect{\dot{P}}(t_r) \right|} } 
\end{align}
The direction of the proper motion vector is converted using
\begin{align}
    \Aboxed{ \murelhatgeotr &=  \frac{ \murelhat - \piE \tE \vect{\dot{P}}(t_r) }{|  \murelhat - \piE \tE \vect{\dot{P}}(t_r) |}}.
    \label{eq:murelhat_conv}
\end{align}
For convenience, we also give this expression in terms of the vector-form microlens parallax:
\begin{align}
\Aboxed{\tEgeotr &= \tE \frac{ \piE }{\left| \piEvec - \piE^2 \tE \vect{\dot{P}}(t_r) \right|} }
\end{align}
The microlensing parallax vector conversion is
\begin{align}
    \piEvecgeotr &= \piE \frac{\murelvecgeotr}{\murelgeotr} \nonumber \\
    &= \piE \frac{\murelvec - \pirel \vect{\dot{P}}(t_r)}{\murelgeotr} \nonumber \\
    % &= \piE \frac{\murelvec}{\murel} \frac{\murel}{\murelgeotr} - \piE \pirel \frac{\vect{\dot{P}}(t_r)}{\murelgeotr} \nonumber \\
    % &=  \frac{\murel}{\murelgeotr} \piEvec - \piE^2 \frac{\thetaE }{\murelgeotr} \vect{\dot{P}}(t_r) \nonumber \\
    % &=  \frac{\tEgeotr}{\tE} \piEvec - \piE^2 \tEgeotr \vect{\dot{P}}(t_r) \nonumber \\
    &=  \frac{\tEgeotr}{\tE} \left( \piEvec - \piE^2 \tE \vect{\dot{P}}(t_r) \right) \nonumber \\
\Aboxed{ \piEvecgeotr &=  \piE \frac{ \piEvec - \piE^2 \tE \vect{\dot{P}}(t_r) }{|  \piEvec - \piE^2 \tE \vect{\dot{P}}(t_r) |}}
\end{align}
and the amplitude $\piE$ is invariant.

The closest approach time in the geocentric($t_r$) frame is calculated by finding when the time derivative of $|\uvecgeotr|$ is zero,
\begin{align}
    \frac{d |\uvecgeotr|^2 }{dt} &= 0.
\end{align}
See Appendix \ref{sec:math_u0} for the full derivation. 
Plugging in and solving for the closest approach time gives,
\begin{align}
\Aboxed{ \tnotgeotr
    = 
    \tnot - \tEgeotr 
   \murelhatgeotr \cdot \left[\uveco - \piE \Pvec(t_r) - (\tnot - t_r) \piE \Pdotvec(t_r) \right]
   }.
\end{align}
For completeness, we restate the closest approach distance
\begin{align}
\Aboxed{
     \uvecogeotr = \uveco 
    + \left( \frac{\tnotgeotr - \tnot}{\tE} \right) \murelhat
    - \piE \Pvec(t_r)
    - (\tnotgeotr - t_r) \piE \Pdotvec(t_r) 
    }
\end{align}
Note that all of the above variables are available in photometry-only fits.

\begin{figure}
    \centering
    \includegraphics[scale=0.4]{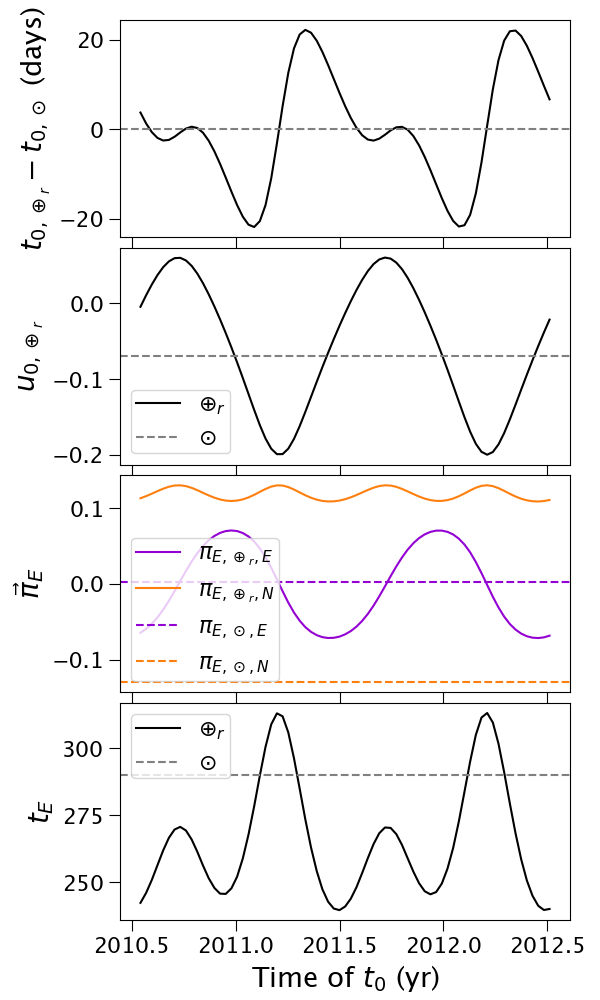}
    \caption{
    Heliocentric ($\odot$) vs. geocentric($t_r$) ($\earth_r$) parameters for an event located towards the Galactic bulge with the following parameters: 
    $\tE = 290.12$ days, $\uo = -0.07$, $\piEvec = $[0.003, -0.13], 
    and variable $\tnot$. 
    For each conversion, the reference time is taken as $\tnot$.
    }
    \label{fig:helio_vs_geo}
\end{figure}

The differences between heliocentric and geocentric($t_r$) can be quite large (Figure \ref{fig:helio_vs_geo}).
BAGLE provides convenience functions to convert between the two coordinate systems in the {\tt frame\_convert} module.

\begin{lstlisting}[language=Python]
from bagle import frame_convert

frame_convert.convert_helio_geo_phot(ra, dec, 
                           t0_in, u0_in, tE_in, 
                           piEE_in, piEN_in, t0par)
\end{lstlisting}

\subsection{Geocentric($t_r$) $\rightarrow$ Heliocentric}

The equations can be reversed to convert from the geocentric($t_r$) parameters to heliocentric parameters.
First, there are a few useful relationships.
\begin{align}
    \murelvec &=  \murelvecgeotr + \pirel \Pdotvec (t_r) 
    \\
    \murelhat &=  \frac{ \murelvecgeotr + \pirel \Pdotvec (t_r) }{
    |\murelvecgeotr + \pirel \Pdotvec (t_r)| }
    \nonumber \\
     &=  \frac{ \murelhatgeotr + \piE \tEgeotr \Pdotvec (t_r) }{
    \left| \murelhatgeotr + \piE \tEgeotr \Pdotvec (t_r) \right| }
\end{align}
The Einstein crossing time conversion is:
\begin{align}
    \tE &= \tEgeotr \frac{\murelgeotr}{\murel} 
    \\
    &=  \tEgeotr \frac{\murelgeotr}{|\murelvecgeotr + \pirel \Pdotvec(t_r) |}   
    \\
    &=  \tEgeotr \frac{1}{|\murelhatgeotr + \piE \tEgeotr \Pdotvec(t_r) |}  
\end{align}

% Closest approach occurs.
% \begin{align}
%     0 &= \murelvec \cdot \uvec(\tnot)  \\
%       &= \murelvec \cdot \uveco \\
    %   &= \murelvec \cdot \left[
    %   \uvecogeotr - \left( \frac{\tnotgeotr - \tnot}{\thetaE} \right) \murelvecgeotr + \piE \Pvec(t_r) 
    %  + (\tnot - t_r) \piE \Pdotvec (t_r) \right]
    %  \\
    %   &= \murelvec \cdot \left[
    %   \uvecogeotr - \left( \frac{\tnotgeotr - \tnot}{\thetaE} \right) \murelvecgeotr + \piE \Pvec(t_r) 
    %  - (\tnotgeotr - \tnot ) \piE \Pdotvec (t_r) 
    %  + (\tnotgeotr - t_r ) \piE \Pdotvec (t_r) 
    %  \right]
% \end{align}
The closest approach time occurs at
\begin{align}
%     \left( \frac{\tnotgeotr - \tnot}{\thetaE} \right) \murelvecgeotr 
%      + (\tnotgeotr - \tnot ) \piE \Pdotvec (t_r)
%     = \murelvec \cdot \left[
%       \uvecogeotr + \piE \Pvec(t_r) 
%      + (\tnotgeotr - t_r ) \piE \Pdotvec (t_r) 
%      \right]
%      \nonumber \\
%      \left( \frac{\tnotgeotr - \tnot}{\tE} \right) 
%      \left[ \murelhatgeotr 
%      +  \piE \tE \Pdotvec (t_r) 
%      \right]
%     = \murelvec \cdot \left[
%       \uvecogeotr + \piE \Pvec(t_r) 
%      + (\tnotgeotr - t_r ) \piE \Pdotvec (t_r) 
%      \right]
%      \nonumber \\
%      \tnot
%     = \tnotgeotr - \tE \frac{ \murelvec \cdot \left[
%       \uvecogeotr + \piE \Pvec(t_r) 
%      + (\tnotgeotr - t_r ) \piE \Pdotvec (t_r) 
%      \right]
%      }{
%      \left[ \murelhatgeotr 
%      +  \piE \tE \Pdotvec (t_r) 
%      \right]
%      } \hspace{1in}
%      \nonumber \\   
%      \tnot
%     = \tnotgeotr - \tE \frac{ \murelhat \cdot \left[
%       \uvecogeotr + \piE \Pvec(t_r) 
%      + (\tnotgeotr - t_r ) \piE \Pdotvec (t_r) 
%      \right]
%      }{
%      \left| \murelhatgeotr 
%      +  \piE \tE \Pdotvec (t_r) 
%      \right|
%      } \hspace{1in}
%      \nonumber \\        
% \Aboxed{     
%      \tnot
%     = \tnotgeotr - \tE \frac{ \left[ \murelvecgeotr + \pirel \Pdotvec(t_r) \right]
%     \cdot \left[
%       \uvecogeotr + \piE \Pvec(t_r) 
%      + (\tnotgeotr - t_r ) \piE \Pdotvec (t_r) 
%      \right]
%      }{
%      \left[ \murelhatgeotr 
%      +  \piE \tE \Pdotvec (t_r) 
%      \right]
%      }
% }  
\Aboxed{     
     \tnot
    = \tnotgeotr - \tE  \murelvec
    \cdot \left[
      \uvecogeotr + \piE \Pvec(t_r) 
     + (\tnotgeotr - t_r ) \piE \Pdotvec (t_r) 
     \right]
}  
\end{align}
The closest approach distance conversion is
\begin{align}
\Aboxed{
     \uveco 
    = \uvecogeotr  
    + \left( \frac{\tnot - \tnotgeotr   }{\tEgeotr} \right) \murelhatgeotr
    + \piE \Pvec(t_r)
    + (\tnot - t_r) \piE \Pdotvec(t_r)
    }
\end{align}

For additional discussion of heliocentric and geocentric microlensing parameter transformations, see \citet{An:2002}, \citet{Gould:2004}, \citet{Jung:2019}.

\begin{widetext}
%\begin{tcolorbox}[width=\textwidth]
\section{Derivation of $\uvecgeotr$ formulation}
\label{sec:math_geotr}
The derivation for Eq.~\ref{ref:eq_eom_geotr} starts with the full equation of motion in the true geocentric frame:
\begin{align}
    \uvecgeo(t) &= \uvec(t) - \piE \vect{P}(t) \nonumber \\   
     &= \uveco + \left( \frac{t - \tnot}{\thetaE} \right) \murelvec - \piE \Pvec (t).
\end{align}
We then convert to the geocentric($t_r$) quantities by performing a rectilinear coordinate transformation and subtracting the Earth-Sun offset in position and velocity at the reference time, $t_r$,
\begin{align}
    \uvecgeotr(t) &= \uvec(t) - \piE \Pvec(t_r) 
    - (t - t_r) \piE \Pdotvec(t_r).
\end{align}
We define the linear proper motion in this new frame as 
\begin{align}
    \murelvecgeotr &= \murelvec - \pirel \Pdotvec (t_r),
\end{align}
such that the linear equation of motion in this frame becomes:
\begin{align}
    \uvecgeotr(t) &= 
    % \uveco 
    % + \left( \frac{t - \tnot}{\thetaE} \right) \murelvec
    % - \piE \Pvec(t_r)
    % - (t - t_r) \piE \Pdotvec(t_r) \\
    %  &= \uveco 
    % + \left( \frac{t - \tnot}{\thetaE} \right) [\murelvecgeo + \pirel \Pdotvec (t_r) ]
    % - \piE \Pvec(t_r)
    % - (t - t_r) \piE \Pdotvec(t_r) \\
     % &= 
     \uveco 
    + \left( \frac{t - \tnot}{\thetaE} \right) \murelvecgeo 
    - \piE \Pvec(t_r)
    - (\tnot - t_r) \piE \Pdotvec(t_r).
\end{align}
The closest approach distance can be defined as
\begin{align}
    \uvecogeotr &= \uvecgeotr(\tnotgeotr) \\
    % &= \uveco 
    % + \left( \frac{\tnotgeotr - \tnot}{\thetaE} \right) \murelvec
    % - \piE \Pvec(t_r)
    % - (\tnotgeotr - t_r) \piE \Pdotvec(t_r) \\
    &= \uveco 
    + \left( \frac{\tnotgeotr - \tnot}{\thetaE} \right) \murelvecgeo 
    - \piE \Pvec(t_r)
    - (\tnot - t_r) \piE \Pdotvec(t_r).
    \label{eq:uo_helio_geotr_conv}
\end{align}
Then we solve for the heliocentric quantities,
\begin{align}
    \murelvec &=  \murelvecgeotr + \pirel \Pdotvec (t_r) 
    \label{eq:murel_geotr_helio_conv} \\
    \uveco &= 
     \uvecogeotr - \left( \frac{\tnotgeotr - \tnot}{\thetaE} \right) \murelvecgeotr + \piE \Pvec(t_r) 
     + (\tnot - t_r) \piE \Pdotvec (t_r)
     \label{eq:uo_geotr_helio_conv}
\end{align}
Substituting these into $\uvecgeo(t)$ yields,
\begin{align}
    \uvecgeo(t) &= 
    % \uvecogeotr - \left( \frac{\tnotgeotr - \tnot}{\thetaE} \right)  
    % \murelvecgeotr + \piE \Pvec(t_r) + (\tnot - t_r) \piE \Pdotvec (t_r) \\
    %     &\hspace{1in} + \left( \frac{t - \tnot}{\thetaE} \right) [\murelvecgeotr + \pirel \vect{\dot{P}} (t_r)] 
    % - \piE \Pvec (t) 
    % \nonumber \\
    %  &= \uvecogeotr + \left( \frac{t - \tnotgeotr}{\thetaE} \right) \murelvecgeotr  
    %  +  \left( t - t_r \right) \piE \Pdotvec (t_r)  + \piE \Pvec (t_r) - \piE \Pvec (t) 
    % \nonumber \\
    %  &= 
     \uvecogeotr + \left( \frac{t - \tnotgeotr}{\thetaE} \right) \murelvecgeotr  
     - \piE \left[  \Pvec(t) - \Pvec (t_r) - \left( t - t_r \right) \Pdotvec (t_r) \right]
\end{align}
\section{Derivation for $\tnotgeotr$}
\label{sec:math_u0}
We solve for the closest approach time in the geocentric($t_r$) frame by finding when the time derivative of $|\uvecgeotr|$ is zero,
\begin{align}
    \frac{d |\uvecgeotr|^2 }{dt} &= 2 \ugeotrE \frac{d \ugeotrE}{dt} + 2 \ugeotrN \frac{d\ugeotrN}{dt} = 0, 
\end{align}
which gives
\begin{align}
    \ugeotrE  \frac{d \ugeotrE}{dt} = - \ugeotrN  \frac{d \ugeotrN}{dt} \\
    \frac{\ugeotrE  \; \murelgeotrE}{\thetaE} = 
      -  \frac{ \ugeotrN \; \murelgeotrN}{\thetaE}
\end{align}
We plug in and solve for $\tnotgeotr$,
\begin{align}
    \murelgeotrE 
    \left[ 
    \uoE 
    + \left( \frac{\tnotgeotr - \tnot}{\thetaE} \right) \murelgeotrE 
    - \piE \PE(t_r)
    - (\tnot - t_r) \piE \PdotE(t_r) 
    \right]
    \nonumber \\
    \hspace{1in} = - \murelgeotrN
    \left[ 
    \uoN 
    + \left( \frac{\tnotgeotr - \tnot}{\thetaE} \right) \murelgeotrN 
    - \piE \PN(t_r)
    - (\tnot - t_r) \piE \PdotN(t_r) 
    \right]
\end{align}
%then
% \begin{gather}
%     \left( \frac{\tnotgeotr - \tnot}{\thetaE} \right) 
%     \left[ \murelgeotrE^2 + \murelgeotrN^2 \right]
%     = 
%     \left( \frac{\tnotgeotr - \tnot}{\tEgeotr} \right) 
%     \murelgeotr
%     = \hspace{2in} 
%     \nonumber \\
%     - \murelgeotrE \left[\uoE - \piE P_E (t_r) 
%     - (\tnot - t_r) \piE \PdotE(t_r) \right] 
%     - \murelgeotrN \left[\uoN - \piE P_N (t_r) 
%     - (\tnot - t_r) \piE \PdotN(t_r) \right]
%     \nonumber
% \end{gather}
% and
% \begin{gather}
%  \frac{\tnotgeotr - \tnot}{\tEgeotr} 
%     = 
%     - \frac{ \murelgeotrN [\uoN - \piE P_N (t_r) 
%                 + (\tnot - t_r) \piE \PdotE(t_r)] +
%              \murelgeotrE [\uoE - \piE P_E (t_r) 
%                 + (\tnot - t_r) \piE \PdotN(t_r)] }{ 
%     \murelgeotr } 
%     \nonumber
% \end{gather}
 and finally
\begin{align}
 \tnotgeotr
    = 
    \tnot - \tEgeotr 
   \murelhatgeotr \cdot \left[\uveco - \piE \Pvec(t_r) - (\tnot - t_r) \piE \Pdotvec(t_r) \right]
\end{align}
%\end{tcolorbox}

\section{Derivation for Astrometric Blending}\label{sec:math_delta_ll}

Here, we calculate the astrometric shift due to microlensing for a luminous source, luminous lens, and blending from neighbors. Figure \ref{fig:astrom_geom} shows the event geometry, where our frame places the lens at the origin. 
\begin{figure}
    \centering
    \includegraphics[width=0.5\linewidth]{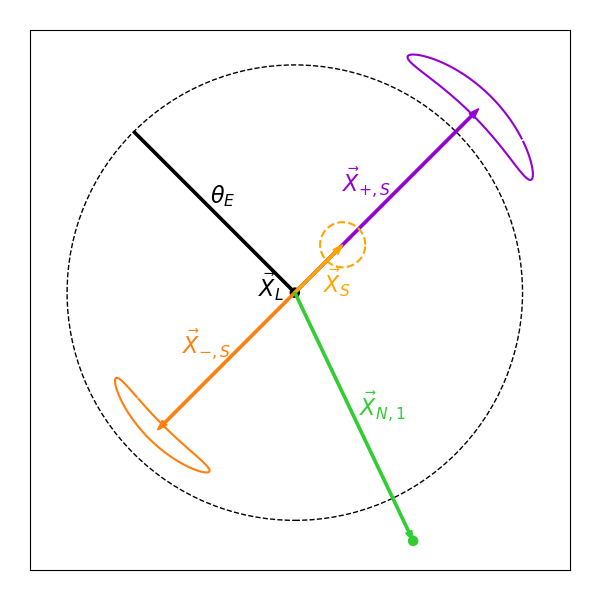}
    \caption{Geometry parameterization for our astrometric shift calculation, with parameters defined in Table \ref{tab:astrom_geom}.}
    \label{fig:astrom_geom}
\end{figure}

\begin{table}[]
    \centering
    \begin{tabular}{|l|l|} 
        \hline
        symbol & definition \\ \hline
        $\vect{X_L}=\vec{0}$ & lens position \\
        $\vect{X_S}$ & true (unlensed) source position \\
        $\vect{X_{N,i}}$ & (i-th) neighbor position \\
        $\vect{X}_{+, S}$ & major image position \\ 
        $\vect{X}_{-, S}$ & minor image position \\ 
        $\theta_E$ & Einstein radius \\ 
        $F_L$ & lens flux \\
        $F_S$ & (baseline) source flux \\
        $F_{N,i}$ & (i-th) neighbor flux \\ 
        $\vec{u} = (\vect{X_S}-\vect{X_L})/\theta_E = \vect{X_S}/\theta_E$ & source-lens separation \\
        $A_1$ & major image magnification factor \\
        $A_2$ & minor image magnification factor \\
        \hline
    \end{tabular}
    \caption{Description of the variables defining our astrometric shift calculation.}
    \label{tab:astrom_geom}
\end{table}

The relevant parameters are defined in Table \ref{tab:astrom_geom}. These parameters involve the solutions to the lens equation, which we can write as:
\begin{equation*}
    \vect{X}_{\pm, S} = \frac{1}{2} \left(1\pm\frac{\sqrt{u^2 + 4}}{u}\right) \vect{X_S}.
\end{equation*}
The magnification of each image is then:
\begin{equation*}
    A_{1,2} = A_{\pm} = \pm\frac{1}{2} + \frac{u^2 + 2}{2u \sqrt{u^2 + 4}}.
\end{equation*}

\noindent The centroid of the unlensed system is
\begin{equation*}
   \vect{X_{cent, unlensed}} = \frac{F_S \vect{X_S} + \sum_i F_{N,i} \vect{X_{N,i}}}{F_S + F_L + \sum_i F_{N,i}}.
\end{equation*}

\noindent The centroid of the lensed system is shown below, in the case where the lens is at the origin ($\vect{X}_L = 0$).
\begin{equation*}
    \vect{X_{cent, lensed}} = \frac{A_1 F_S \vect{X}_{+, S} + A_2 F_S \vect{X}_{-, S} + \sum_i F_{N,i} \vect{X_{N,i}}}{A_1 F_S + A_2 F_S + F_L + \sum_i F_{N,i}}.
\end{equation*}

\noindent Thus, the observed centroid shift $\vect{\delta_{c, obs}}$ is 
\begin{equation*}
    \vect{\delta_{c, obs}} = \vect{X_{cent, unlensed}} - \vect{X_{cent, lensed}} = \frac{A_1 F_S \vect{X}_{+, S} + A_2 F_S \vect{X}_{-, S} + \sum_i F_{N,i} \vect{X_{N,i}}}{A_1 F_S + A_2 F_S + F_L + \sum_i F_{N,i}} - \frac{F_S \vect{X_S} + \sum_i F_{N,i} \vect{X_{N,i}}}{F_S + F_L + \sum_i F_{N,i}}.
\end{equation*}

\noindent We can re-parameterize the source position in terms of $u$. For the numerator, the directional vectors matter: 
\begin{equation*}
    A_1 \vect{X}_{+, S} + A_2 \vect{X}_{-, S} = \frac{u^2 + 3}{\sqrt{u^2 + 4}} \theta_E \hat{\vect{X_S}} = \frac{u^2 + 3}{u\sqrt{u^2 + 4}} \vect{X_S}.
\end{equation*}

\noindent For the total flux, we only need the total magnification,
\begin{equation*}
    A_1 + A_2 = \frac{u^2 + 2}{u\sqrt{u^2 + 4}}.
\end{equation*}

\noindent Thus we can rewrite the observed centroid shift $\vect{\delta_{c, obs}}$ as
\begin{equation} \label{eq:cent_shift}
    \vect{\delta_{c, obs}} = \frac{\frac{u^2 + 3}{u\sqrt{u^2 + 4}} F_S \vect{X_S}+ \sum_i F_{N,i} \vect{X_{N,i}}}{\frac{u^2 + 2}{u\sqrt{u^2 + 4}} F_S + F_L + \sum_i F_{N,i}} - \frac{F_S \vect{X_S} + \sum_i F_{N,i} \vect{X_{N,i}}}{F_S + F_L +  \sum_i F_{N,i}}.
\end{equation}

\noindent Next, we examine this result in some limiting cases.

\subsection{Limit: Dark Lens, No Neighbors}
For a dark lens and no luminous, blended neighbors
($F_L = F_N = 0$), Equation \ref{eq:cent_shift} becomes:
\begin{align*}
    \vect{\delta_{c,S}} &= \frac{\frac{u^2 + 3}{u\sqrt{u^2 + 4}} F_S \vect{X_S} }{\frac{u^2 + 2}{u\sqrt{u^2 + 4}} F_S} - \frac{F_S \vect{X_S} }{F_S} \\
    %&= \Bigg[ \frac{u^2 + 3}{u^2 + 2} - 1 \Bigg] \vect{X_S} \\
    &= \frac{1}{u^2 + 2} \vect{X_S} . % \\
    %&= \frac{\vec{u}}{u^2 + 2} \theta_E,
\end{align*}
The resulting magnitude of the shift,
\begin{equation*}
       \delta_{c,S} = \frac{u}{u^2 + 2} \theta_E,
\end{equation*}
is in agreement with Equation (15) from \citet{Dominik2000}.

\subsection{Limit: Luminous Lens, No Neighbors}
 Next, we consider a luminous lens but no luminous neighbors ($F_N = 0$). We will define $g = F_L / F_S$. Thus:
\begin{align*}
    \vect{\delta_{c, LL}} &= \frac{\frac{u^2 + 3}{u\sqrt{u^2 + 4}} F_S \vect{X_S} }{\frac{u^2 + 2}{u\sqrt{u^2 + 4}} F_S + F_L} - \frac{F_S \vect{X_S} }{F_S + F_L} \\
    &= \Bigg[ \frac{\frac{u^2 + 3}{u\sqrt{u^2 + 4}}}{\frac{u^2 + 2}{u\sqrt{u^2 + 4}} + g} - \frac{1}{1 + g} \Bigg] \vect{X_S} \\
    %&= \Bigg[ \frac{u^2 + 3}{u^2 + 2 + gu \sqrt{u^2 + 4}} - \frac{1}{1+g} \Bigg] \vect{X_S} \\
    %&= \Bigg[ \frac{(u^2 + 3)(1 + g) - (u^2 + 2 + gu \sqrt{u^2 + 4})}{(1 + g)(u^2 + 2 + gu \sqrt{u^2 + 4})} \Bigg] \vect{X_S} \\
    &= \frac{1 + g(u^2 + 3 - u\sqrt{u^2 + 4})}{(1 + g)(u^2 + 2 + gu \sqrt{u^2 + 4})} \vect{X_S} . %\\
    %&= \frac{1 + g(u^2 + 3 - u\sqrt{u^2 + 4})}{(1 + g)(u^2 + 2 + gu \sqrt{u^2 + 4})} \theta_E \vec{u}.
\end{align*}
Then the magnitude of the astrometric shift with a luminous lens is
\begin{equation*}
    \delta_{c,LL} = \frac{u}{1 + g} \frac{1 + g(u^2 - u \sqrt{u^2 + 4} + 3)}{u^2 + 2 + gu\sqrt{u^2 + 4}} \theta_E,
\end{equation*}
which is in agreement with Equation (98) from \citet{Dominik2000}.

\end{widetext}

\bibliographystyle{aasjournalv7}
\bibliography{main.bib}

\end{document}